\documentstyle[epsfig,12pt,preprint,tighten,aps]{revtex}

\begin{document}

\draft

\title{
\rightline{{\tt August 2000}}
\
\\ Comment on ``Neutrino oscillations in the early universe: how
can large lepton asymmetry
 be generated?" }
\author{P. Di Bari$^{1,2}$, R. Foot$^1$,
R. R. Volkas$^1$ and Y. Y. Y. Wong$^1$} \maketitle
\begin{center}
{\small \it $^1$ School of Physics \\ Research Centre for High
Energy Physics\\ The University of Melbourne Vic 3010
\\ Australia\\ $^2$ Istituto Nazionale di Fisica Nucleare (INFN)\\
(dibari, foot, r.volkas, ywong@physics.unimelb.edu.au)}
\end{center}

\vspace{-0.5cm}

\begin{abstract}
We comment on the recent paper by A. D. Dolgov, S. H. Hansen, S.
Pastor and D. V. Semikoz (DHPS) [Astropart.\ Phys.\ {\bf 14}, 79
(2000)] on the generation of neutrino asymmetries from
active--sterile neutrino oscillations. We demonstrate that the
approximate asymmetry evolution equation obtained therein is an
expansion, up to a minor discrepancy,
 of the well-established static approximation equation,
  valid only when the supposedly new higher order correction term is small.
 In the regime where this so-called ``back-reaction'' term is
large and artificially terminates the asymmetry growth, their
evolution equation ceases to be a faithful approximation to the
Quantum Kinetic Equations (QKEs) simply because pure
Mikheyev--Smirnov--Wolfenstein (MSW) transitions have been
neglected. At low temperatures the MSW effect is the dominant
asymmetry amplifier. Neither the static nor the DHPS approach
contains this important physics. Therefore we conclude that the
DHPS results have sufficient veracity at the onset of
explosive asymmetry generation, but are invalid in the ensuing low
temperature epoch where MSW conversions are able to enhance the
asymmetry to values of order $0.2 - 0.37$. DHPS do claim
to find a significant final asymmetry for very large
$\delta m^2$ values. However, for this regime the effective
potential they employed is not valid.
\end{abstract}

%\newpage

\section{Introduction and overview}
\label{sec1}

 Active--sterile neutrino oscillations in the early
universe can generate large differences in the number densities of
relic neutrinos and antineutrinos \cite{ftv,fv1,qke}.\footnote{For
the historical record, the reader should note that Ref.\cite{bd}
contains the claim that the neutrino asymmetry is always small
(less than $10^{-7}$) and hence unimportant for all parameters of
interest. This contrasts starkly with the later studies
\cite{ftv,fv1,qke}.} The resulting ``lepton asymmetries'' have
important phenomenological ramifications: the suppression of
sterile neutrino production because of large matter effects
\cite{ftv,fv1}, a modification to primordial Helium synthesis if a
sufficiently large asymmetry is created for the electron neutrino
prior to or during the Big Bang Nucleosynthesis epoch \cite{qke},
with also the possibility to describe inhomogeneous BBN scenarios
\cite{domains}.

We focus in this Comment on two flavour active--sterile
oscillations. Furthermore we restrict our discussion to that
region of oscillation parameter space which induces large lepton
asymmetry creation well before neutrino decoupling at $T \simeq 1$
MeV. An important issue is the ``final value of the asymmetry'',
that is, the steady state value attained at the end of the
dynamical evolution. The asymmetry is conveniently defined to be
\begin{equation}
L_{\nu_{\alpha}} = \frac{n_{\nu_{\alpha}} -
n_{\overline{\nu}_{\alpha}}}{n_{\gamma}}, \label{def}
\end{equation}
where $n_\psi$ is the number density of species $\psi$, and
$\alpha$ is one of $e$, $\mu$ or $\tau$. In Ref.\ \cite{qke} it is
argued that the final value for $L_{\nu_{\alpha}}$ is of order
$0.2 - 0.37$ for typical parameter choices. This claim is recently
disputed by A. D. Dolgov, S. H. Hansen, S. Pastor and D. V.
Semikoz (DHPS) in Ref.\ \cite{dolgov}, who conclude that although
the asymmetry can rise by five orders of magnitude above the
baryon asymmetry level of $\sim 10^{-10}$, it cannot reach the
$0.2 - 0.37$ range found in Ref.\ \cite{qke}. The purpose of this
Comment is to explain the flaws in the DHPS analysis.

The basic reason turns out to be very simple: DHPS neglect the
pure Mikheyev--Smirnov--Wolfenstein (MSW) effect \cite{msw}. As
 explained in Ref.\cite{qke}, pure MSW transitions
dominate lepton asymmetry evolution at lower temperatures, after
the period of initial explosive $L_{\nu_{\alpha}}$ growth. The
approximate evolution equations derived by DHPS have sufficient
validity to describe the onset of asymmetry growth, but they are
not a good approximation after the explosive amplification phase
ends. Their oversight of the role played by MSW transitions also
leads them to exaggerate the importance of a ``back-reaction
term'' [$B_1$ in Eq.\ (51) of DHPS]. As we will explain, when this
term is large, the MSW effect is also dominant so that their Eq.\
(51) is no longer valid.  The size of the back-reaction is thus a
moot point.

We now clarify the structure of our paper, while simultaneously
providing a  summary of the main points to be made. These remarks
will be of interest to those readers who want to know the gist of
our argument but not the technical details.
\begin{itemize}
\item Section \ref{sec2} contains a quick survey of the Quantum Kinetic
Equations (QKEs) governing active--sterile neutrino evolution
\cite{stod,stod2,bm}. These ``almost exact'' equations are not at
issue; the debate is on their correct solutions. The QKEs
simultaneously incorporate coherent matter-affected evolution,
decoherence due to non-forward neutrino scattering, and
repopulation of active neutrino distribution functions from the
background plasma.
\item Section \ref{sec3} reviews how the essential physics encoded in
the QKEs can be revealed. A fundamental issue is
collision-dominated versus coherent matter-affected
oscillation-driven amplification. The former pertains to
conditions prior to and during the onset of asymmetry growth, in
which case the evolution is well described by the static
approximation of Refs.\ \cite{ftv,fv1}, and by the closely related
adiabatic limit approximation developed in Refs.\
\cite{bvw,new1,new2}. Equation (51) of DHPS is identical to the
leading order term of the static approximation if the $B_1$ term
is neglected, and is thus able to correctly predict the existence
of a critical temperature \cite{ftv} at which asymmetry evolution
enters a brief explosive growth phase. Subsequent evolution,
however, ceases to be collision dominated. In addition, a
significant asymmetry, typically $\sim 10^{-5}$ when explosive
growth stops, now exists in the plasma; the antisymmetric
Wolfenstein term in the matter potential is sufficiently large to
modify the neutrino and antineutrino oscillation patterns in
noticeably dissimilar ways. A vitally important consequence of
this is the separation of the neutrino and antineutrino MSW
resonance momenta \cite{qke}. Depending on the sign of
$L_{\nu_{\alpha}}$, one resonance momentum remains in the body of
the (almost) Fermi--Dirac distribution, while the other is rapidly
relegated to the tail. Figures 1b and 2b depict this phenomenon
for two different oscillation parameter choices. The separation
means that the MSW effect, now dominant over the collision
mechanism, has very asymmetric consequences for neutrinos and
antineutrinos \cite{qke}. For $L_{\nu_{\alpha}}>0$, the resonance
in the body is able to efficiently convert
$\overline{\nu}_{\alpha}$'s  into $\overline{\nu}_s$'s, while the
other is impotent in the scarcely populated tail.  This important
physics was not understood by DHPS. Indeed, when collisions are
absent, neutrino density matrix evolution is driven only by
matter-affected propagation (and the expansion of the universe of
course). Equation (51) of DHPS {\it does not reduce} to the
coherent matter-affected evolution equations, so it does not
contain the pure MSW effect \cite{fv1}. The static approximation,
which yields $\frac{d L_{\nu_{\alpha}}}{d t} = 0$ for zero
collision rate, also neglects the MSW effect. Thus these equations
are valid only in the collision dominated regime. Figures 1a and
2a illustrate our arguments. The solid and dash-dotted lines
represent respectively numerical solutions to the exact
QKEs\footnote{DHPS are unable to successfully integrate the QKEs,
and do not accept these curves as accurate, although the numerical
procedures used, and their stability, are thoroughly discussed in
Ref.\ \cite{ropa}.} and the static approximation equations.
Clearly, the latter describes the explosive growth phase
excellently, but underestimates the subsequent growth of
$L_{\nu_{\alpha}}$. At steady state, the discrepancy is
significant. The dashed curve reveals the physics behind the QKEs
in a very amusing way. The generating code  compares the
decoherence or interaction length $\ell_{\rm int}$ with the
matter-affected oscillation length $\ell_{m}$ at every time step,
with both lengths evaluated at resonance. If $r|_{\rm res} \equiv
(\ell_{\rm int}/\ell_{m})|_{\rm res}$ is  less than one, the
static approximation equations are used to forward the system  in
time. For $r|_{\rm res} > 1$, pure adiabatic MSW evolution is
employed. The quantity $r|_{\rm res}$ therefore measures the
relative importance of collisions and MSW transitions. In the
event $r|_{\rm res} \ll 1$, the MSW effect is damped since
collisions seriously disrupt the coherence. The opposite case has
$r|_{\rm res} \gg 1$, implying that collisions are relatively
negligible. The dashed curve in Fig.\ 1a agrees with the QKEs in
both the high and low $T$ regimes extremely well. There is a
slight discrepancy immediately after explosive growth, since
neither effect dominates at this stage --- the very visible bump
in Fig.\ 2a is an artifact of this ``either/or'' approximation in
this intermediate regime.
\item The DHPS equations underestimate the asymmetry after explosive growth
 even more seriously than does the static approximation, because
of their $B_1$ back-reaction term. Section \ref{sec4} compares
Eq.\ (51) of DHPS with the static approximation. As already noted,
and acknowledged by DHPS, these equations agree to leading order.
We show, however, that the $B_1$ term  is large only when the MSW
effect has taken over as the dominant driver of asymmetry
evolution. Since DHPS Eq.\ (51) and the static approximation both
neglect this effect, they should not be used when MSW transitions
are important. (We will also identify where  the MSW effect is
unwittingly removed in their derivation.) Furthermore, the
expansion parameter adopted by DHPS in  their approximation scheme
is not in fact the correct choice at lower temperatures when there
is a significant asymmetry in the plasma. This misunderstanding
led DHPS to erroneously claim validity for their equations at
later stages in the evolution, especially the role of the
back-reaction term which is but another casualty following from  a
faulty expansion parameter.  In fact, the dramatic termination of
the asymmetry growth due to this bogus term has its origin in the
expression $(1 + x)^{-1} \simeq 1 - x$. The left hand side
corresponds  to the static approximation (with a minor
discrepancy), while the right hand side denotes the DHPS result,
with $x$ identified as the $B_1$ term. The artificially strong
cut-off  arises because the right hand side is used even when $x
\sim 1$.
\item Section \ref{sec5} contains comments on other less important errors
and some misleading statements made by DHPS, while Sec.\
\ref{sec6} is a conclusion.
\item The Appendix provides a translation of DHPS's notation into a more
common {\it lingua franca}  used by other authors. The rather
obscure nomenclature of DHPS proves to be the biggest obstacle to
understanding their analysis.

\end{itemize}

\section{Short review of the Quantum Kinetic Equations}
\label{sec2}

 We consider the case of active--sterile two
state mixing where the weak eigenstates $\nu_{\alpha}$ ($\alpha =
e, \mu$ or $\tau$) and $\nu_{s}$ are linear combinations of two
mass eigenstates $\nu_a$ and $\nu_b$,
\begin{equation}
\nu_{\alpha} = \cos\theta_0 \nu_a + \sin\theta_0 \nu_b,\quad
\nu_{s} = - \sin\theta_0 \nu_a + \cos\theta_0 \nu_b, \label{zwig}
\end{equation}
with $\theta_0$ as the vacuum mixing angle, and $\cos2 \theta_0
> 0$ by definition.  The compact notation
\begin{equation}
s\equiv \sin 2\theta_0,\quad c\equiv \cos 2\theta_0,
\end{equation}
and the convention $\delta m^2 \equiv m^2_b - m^2_a$ will be used
from now on.

The one-body reduced density matrix \cite{stod,stod2,bm} for a
$\nu_{\alpha} \leftrightarrow \nu_s$ system of momentum $p$
 in the early universe
can be parameterised through
\begin{equation}
\rho(y) = {1 \over 2} [P_0(y) + {\bf P}(y)\cdot {\bf \sigma}],
\label{kdf}
\end{equation}
in which ${\bf P}(y) = P_x(y){\bf \hat x} + P_y(y) {\bf \hat y} +
P_z(y){\bf \hat z}$ is the ``polarisation vector'', and ${\bf
\sigma} = \sigma_x {\bf \hat x} + \sigma_y {\bf \hat y} + \sigma_z
{\bf \hat z}$, where $\sigma_i$ are the Pauli matrices. The
quantity $y$ is a dimensionless momentum variable that does not
red-shift by definition:
\begin{equation}
y \equiv \frac{p}{T} \propto  p {\cal R}(t),
\end{equation}
where ${\cal R}$ is the Friedmann--Robertson--Walker scale factor
and $T$ is the temperature. It is understood that the density
matrices and the variables $P_i(y)$ depend also on time $t$ or,
equivalently, temperature $T$, which, for $m_e \lesssim T \lesssim
m_\mu$, are related through $dt/dT = -1/({\cal H} T) \simeq
-M_P/5.44T^3$, where ${\cal H} = \dot{{\cal R}}/{\cal R}$ is the
Hubble parameter and $M_P \simeq 1.22 \times 10^{22}$ MeV is the
Planck mass.

We normalise the density matrices to give momentum distribution
functions
\begin{equation}
f_{\nu_{\alpha}}(y) = {1 \over 2}[P_0(y) + P_z(y)]f^{0}_{\rm
eq}(y), \quad \quad f_{\nu_{s}}(y) = {1 \over 2}[P_0(y) -
P_z(y)]f^{0}_{\rm eq}(y), \label{c}
\end{equation}
for $\nu_{\alpha}$ and $\nu_{s}$ respectively.  The reference
distribution $f^{0}_{\rm eq}(y)$ is of the Fermi--Dirac form [see
Eq.\ (\ref{FDdistribution})] with zero chemical potential. Similar
expressions pertain to antineutrinos, with $P_i(y) \to \overline{
P}_i(y)$. The number densities are obtained from the distribution
functions through
\begin{equation}
n_\psi = \frac{T^3}{2\pi^2} \int_0^{\infty} f_\psi(y) y^2 dy,
\end{equation}
where $\psi$ denotes the species.

The evolution equations for ${\bf P}(y)$ and $P_0(y)$
 are \cite{stod,stod2,bm,bvw}
\begin{eqnarray}
\label{eq:b1} {\partial {\bf P}(y) \over \partial t} & = & {\bf
V}(y) \times {\bf P}(y) - D(y)[P_x (y) {\bf \hat x} + P_y (y){\bf
\hat y}]+ {\partial P_0(y) \over \partial t}\, {\bf \hat z},
\nonumber \\ {\partial P_0(y) \over \partial t} & \simeq &
\Gamma(y)\left\{{f_{\rm eq}(y) \over f^0_{\rm eq}(y)} - {1\over
2}[P_0 (y) + P_z (y)] \right\}.
\end{eqnarray}
The damping or decoherence function $D$ is related to the  total
collision rate for $\nu_{\alpha}$ of momentum $y$ with the
background plasma, $\Gamma(y)$, via \cite{stod2,bvw,ekt}
\begin{equation}
D(y) = \frac{\Gamma(y)}{2} \simeq
\frac{1}{2}k_{\alpha}\,G_F^2T^5\,y,
\end{equation}
with $k_{e} \simeq 1.27$ and $k_{\mu}=k_{\tau} \simeq 0.92$ (for
$m_e \lesssim T \lesssim m_\mu$).\footnote{These values neglect
the Pauli blocking factors. Including Pauli blocking factors
lead corrections of order $10\%$ for these quantities\cite{bvw,dolgov}.}
It is useful to define the
quantity
\begin{equation}
{\ell}_{\rm int}\equiv D^{-1},
\end{equation}
to be identified as the decoherence or interaction length.

The function $f_{\rm eq}(y)$ is the Fermi--Dirac distribution,
\begin{equation}
\label{FDdistribution} f_{\rm eq}(y) \equiv {1 \over 1 + e^{y-
\tilde{\mu}_{\alpha}}},
\end{equation}
where $\tilde{\mu}_{\alpha} \equiv \mu_{\nu_\alpha}/T$ (and
$\tilde{\mu}_{\overline{\alpha}} \equiv
\mu_{\overline{\nu}_{\alpha}}/T$ for antineutrinos) is a
dimensionless chemical potential which does not red-shift but
depends on the neutrino asymmetry.

The  matter potential vector \cite{stod2,bm}, ${\bf V}(y) = V_x
(y) {\bf \hat x} + V_z (y) {\bf \hat z}$, has coefficients
\cite{msw,nr}
\begin{equation}
V_x = \frac{\sin 2\theta_0}{\ell_0},\quad
V_z = -\frac{\cos2\theta_0}{\ell_0} + V_{\alpha},
\label{sf}
\end{equation}
where $\ell_0 \equiv 2yT/\delta m^2$ is the vacuum oscillation
length. The quantity  $V_{\alpha}(y)$ is the effective potential
which takes the form
\begin{equation}\label{eq:effp}
V_{\alpha}(y) = \pm \sqrt{2}G_F n_{\gamma}L^{(\alpha)}-
            \sqrt{2} G_F n_{\gamma}{A_{\alpha} T^2 y\over
            M^2_W},
\end{equation}
where the $+(-)$ sign corresponds to neutrino (antineutrino)
ensemble, $G_F$ is the Fermi constant, $M_W$ is the $W$-boson
mass, $A_e \simeq 17$ and $A_{\mu, \tau} \simeq 4.9$ (for $m_e
\lesssim T \lesssim m_\mu$), and $L^{(\alpha)}$ is the {\it
effective total lepton number} (for the $\alpha$-neutrino
species).  This last quantity is essentially a sum of the
individual asymmetries:
\begin{equation}
L^{(\alpha)} \equiv L_{\nu_\alpha} + L_{\nu_e} + L_{\nu_\mu}
+ L_{\nu_\tau} + \eta,
\end{equation}
with $\eta$ being a small term due to the asymmetry of the
electrons and nucleons --- typically of order  $|\eta| \sim
5\times 10^{-10}$, calculated from \cite{nr}
\begin{equation}
\eta = L_e -{1 \over 2}L_N\; \; ({\rm for}\; \alpha=e),\quad \eta
= - {1\over 2}L_N\; \; ({\rm for}\; \alpha=\mu,\tau).
\end{equation}
For simplicity we shall henceforth write $L$ instead of
$L^{(\alpha)}$.

We now introduce $v_\alpha$ and the dimensionless variables $a$ and $b$ by
\begin{eqnarray}\label{eq:vtal}
v_{\alpha} &\equiv &{\ell}_0\,V_{\alpha}=\mp a + b, \nonumber \\
 a &\equiv& - \sqrt{2}G_F n_{\gamma}L {\ell}_0
\simeq - 8.033 L y \left({{\rm eV}^2 \over \delta m^2}\right)
\left({T \over {\rm MeV}}\right)^4, \nonumber\\ b &\equiv&
-\sqrt{2} G_{F} n_{\gamma} {\ell}_{0}\, \frac{A_{\alpha}T^{2}
y}{M^{2}_{W}} = - y^2 \left(\frac{T}{T^{\alpha}}\right)^{6}\,
\left(\frac{{\rm eV}^{2}}{\delta m^{2}}\right),
\end{eqnarray}
where, in the definition of $v_{\alpha}$, $-(+)$ pertains to
neutrinos (antineutrinos), and
\begin{equation}
T^{\alpha}\simeq 18.9\, (23.4)\,{\rm MeV},\quad {\rm for}\;
\alpha=e \,(\mu,\tau),
\end{equation}
is a convenient reference temperature.  In this new notation, the
matter-affected oscillation length is given by
\begin{equation}
{\ell}_m \equiv \frac{1}{\sqrt{V_x^2 + V_z^2}} = {{\ell}_0\over
\sqrt{ s^2 + \left(c-v_{\alpha}\right)^2}}.
\end{equation}
We will also have cause to consider the ratio
\begin{equation}
d \equiv \frac{\ell_{0}}{{\ell}_{\rm int}}= k_{\alpha} G^{2}_{F}
\frac{T^{6} y^{2}}{\delta m^{2}}=\delta b
\end{equation}
where $\delta \simeq 0.8\, (2.0) \times 10^{-2}$ for $\alpha =e\,
(\mu,\tau)$. In addition, a very important role is played by
\begin{equation}
r \equiv \frac{{\ell}_{\rm int}}{{\ell}_{m}} = \frac{\sqrt{V_x^2 +
V_z^2}}{D} = \frac{\sqrt{s^{2}+(c-v_{\alpha})^{2}}}{d},
\label{lintoverlm}
\end{equation}
which is the ratio of the interaction and matter-affected oscillation lengths.
This quantity parameterises the relative importance of decohering
collisions and coherent oscillations on the evolution of the density matrices.

Note that the MSW resonance condition is given by $V_z = 0
\Rightarrow  c=v_{\alpha}=\mp a + b$, where $-\, (+)$ is used for
neutrinos (antineutrinos). Evaluating $r$ at resonance,
\begin{equation}
r|_{\rm res} = \left. \frac{V_x}{D}\right|_{\rm res}= \left.
\frac{s}{d}\right|_{\rm res}, \label{rimatres}
\end{equation}
we find that the relative importance of collisions and coherent
oscillations is strongly  dependent on $s \equiv \sin 2\theta_0$.

To make our argument easier to follow, it is useful to distinguish
two different asymptotic regimes for the resonance condition.

\medskip
 \noindent 1.\
{\it Negligible total lepton number} $|a|\ll |b|, c$. This is the
situation prior to the explosive growth phase, in which the
resonance condition
\begin{equation}
c = b,
\label{cequalsb}
\end{equation}
 is identical for neutrinos and antineutrinos.
Equation (\ref{cequalsb}) can be recast in the form
\begin{equation}\label{eq:tres}
T_{\rm res}^{0}(y)\equiv T^{\alpha} \left(\frac{|\delta
m^{2}|}{{\rm eV}^{2}}\right)^{\frac{1}{6}}
y^{-\frac{1}{3}}c^{\frac{1}{6}},
\end{equation}
where $T_{\rm res}^0$ is the temperature at which neutrinos or
antineutrinos of momentum $y$ are at resonance. For the body of
the Fermi--Dirac distribution, i.e., $0.1 \leq y \leq 10$, one can
then deduce from this expression the temperature range in which
oscillations will be most significant. Note, however, that since
Eq.\ (\ref{eq:effp}) holds only for $T \lesssim 150\,{\rm MeV}$,
we are restricted to the regime $|\delta m^{2}|/{\rm
eV}^{2}\lesssim10^{4}$ for $c \sim 1$.

Using the definition of $T_{\rm res}^0$ one may rewrite the
function $b$ via
\begin{equation}\label{eq:b}
|b|= c \left({T\over T_{\rm res}^0}\right)^6.
\end{equation}
This will be useful later. Equation (\ref{cequalsb}) can also be
rewritten in the form
\begin{equation}\label{eq:yres}
y_{\rm res}^{0} = \left(\frac{T^{\alpha}}{T}\right)^{3}
\left(\frac{|\delta m^{2}|}{{\rm eV}^{2}}\right)^{\frac{1}{2}}
c^{\frac{1}{2}},
\end{equation}
which shows that the resonance momentum moves from low to high
values as the temperature decreases. At $y=y^0_{\rm res}$, the
condition $|a| \ll |b|=c$
 is equivalent to
\begin{equation}\label{eq:Lt}
L \ll   1.0\,(0.4)\times 10^{-6} \left({|\delta
m^2|\over {\rm eV}^2}\right)^{\frac{1}{3}} (y_{\rm res}^0)^{1\over
3} c^{\frac{1}{3}} \equiv L_t,
\end{equation}
for $\alpha=e\,(\mu, \tau)$. The most relevant choice for $y_{\rm
res}^0$ is about 2, because asymmetry growth starts approximately
at this point as the resonance momentum moves form low to high
values with decreasing $T$. This value is approximately correct if
sterile neutrino production has been negligible until that moment
(which is true for small enough mixing angles).

If $L\gg L_{t}$, the Wolfenstein term $a$ in the effective
potential becomes dominant and two different resonances exist, one
on either side of $y_{\rm res}^0$ in the momentum distribution. If
$L$ is positive, the resonance at the lower (higher) value occurs
only for antineutrinos (neutrinos), and vice versa if $L$ is
negative. This ``separation of the resonances'' is crucial for
understanding the mechanism of neutrino asymmetry generation.

\medskip \noindent 2.\ {\it Negligible finite-temperature term}
$|b|\ll |a|, c$. This situation holds after a substantial $L$ has
been created. The resonance condition is now approximately given
by
\begin{equation}
 c = \mp a,
\label{cequalspma}
\end{equation}
where $-(+)$ refers to neutrinos (antineutrinos). In the case
$L>0$, the resonance occurs for antineutrinos (neutrinos) if
$\delta m^{2}<0$ ($\delta m^{2}>0$). The opposite holds if $L<0$.
It is important to recall that the asymmetry generation mechanism
actually requires $\delta m^2 < 0$.

Equation (\ref{cequalspma}) implies that the resonance momentum takes the value
\begin{equation}\label{eq:yreslow}
y^{\rm low}_{\rm res} = y^{0}_{\rm res}
\left[\frac{L_{t}}{|L|}\right]^{\frac{3}{4}} \simeq {c\over 8.033}
\frac{|\delta m^{2}|}{{\rm eV}^{2}} \frac{1}{L (T/{\rm MeV})^{4}}.
\end{equation}
Notice that $y^{\rm low}_{\rm res} < y^{0}_{\rm res}$ when $|L| > L_t$.

\section{Understanding the lepton asymmetry evolution curves}
\label{sec3}

The Quantum Kinetic Equations can be numerically integrated
and lepton asymmetry
evolution thereby extracted. In terms of the density matrices, the asymmetry
is given by
\begin{equation}
L_{\nu_{\alpha}} = \frac{1}{2 n_{\gamma}} \frac{T^3}{2 \pi^2}
\int_0^{\infty} \left[ P_0(y) + P_z(y) - \overline{P}_0(y) -
\overline{P}_z(y) \right] f^0_{\rm eq}(y) y^2 dy. \label{LfromPs}
\end{equation}
Taking the time derivative of Eq.\ (\ref{LfromPs}), and imposing
$\alpha + s$ lepton charge conservation, i.e.,
\begin{equation}
\label{conservation}
\int^{\infty}_0
\left[
\frac{\partial P_0(y)}{\partial t}
-\frac{\partial \overline{P}_0(y)}{\partial t} \right]
f^0_{\rm eq}(y) y^2 dy  \propto
\frac{\partial}{\partial t} \left(\frac{
n_{\nu_{\alpha}} + n_{\nu_s} - n_{\overline{\nu}_{\alpha}}
- n_{\overline{\nu}_s}}{n_{\gamma}} \right)
= 0,
\end{equation}
 one easily
deduces that
\begin{equation}
\frac{d L_{\nu_{\alpha}}}{dt} = \frac{1}{2 n_{\gamma}}
\frac{T^3}{2 \pi^2}\int_0^{\infty} V_x (P_y - \overline{P}_y)
f^0_{\rm eq}y^2 dy. \label{formaldLdt}
\end{equation}
In principle, one could use Eq.\ (\ref{LfromPs}) to directly
calculate the asymmetry; Equation (\ref{formaldLdt}) is formally
redundant.  However, for very small asymmetries, the right hand
side of the former equation is the difference of two
numbers which are large relative to $L_{\nu_{\alpha}}$.
Equation (\ref{formaldLdt}) should therefore be used to
circumvent the numerical stability problems potentially posed by
the direct procedure.

DHPS were apparently unable to achieve successful numerical
integration of the QKEs for 
$|\delta m^2| > 10^{-7}\ eV^2$. We would guess that at least part of the
reason for this was that they did not use Eq.\ (\ref{formaldLdt}).

The solid lines in Figs.\ 1a and 2a show our numerically stable
asymmetry evolution curves from integrating the QKEs.  There are
four phases in the evolution: Above the critical temperature
$T_c$, the asymmetry $L_{\nu_{\alpha}}$
evolves from its small initial value to the
$\eta \sim 10^{-10}$ level such that $L \ll \eta$.
At $T = T_c$, explosive growth occurs for a
short period, during which the asymmetry rises to roughly $10^{-5}$
for the typical choice $\delta m^2 \sim 1$ eV$^2$.
After this phase, asymmetry growth
continues at a more leisurely power law rate. One can measure from
the graphs that $L_{\nu_{\alpha}} \propto T^{-4}$. Further
evolution brings the asymmetry to a steady state value in the
range $0.2 - 0.37$ depending on the oscillation parameters chosen.

The post explosive growth behaviour  is disputed by DHPS. In this
section, we review the physical explanation for the
$L_{\nu_{\alpha}}$ curves obtained by solving the QKEs, and
demonstrate concordance between these curves and those generated
by certain approximate evolution equations which highlight the
essential physical processes occurring in the various regimes.

\subsection{Collision dominated evolution and the static approximation}

The neutrino collision rate $\Gamma$ increases as $T^5$. We
expect, therefore, that at high enough temperatures, collisions
are sufficiently frequent to seriously disrupt the coherent
matter-affected oscillatory evolution of the ensemble. The finite
temperature $b$ term in the effective potential also scales as
$T^5$, and predominates at high $T$ if the initial neutrino
asymmetry is small (say of the order of the baryon asymmetry). We
focus first on this regime.

 A useful and
intuitively reasonable measure of the importance of collisional
decoherence is afforded by the quantity $\left. r \right|_{\rm
res} \equiv \left. ({\ell}_{\rm int}/{\ell}_{m})\right|_{\rm res}$
defined in Eq.\ (\ref{rimatres}). The collision dominated regime
then corresponds to $\left. r \right|_{\rm res} \ll 1$. In this
regime, one can develop a physical picture for distribution
function evolution that yields approximate time-development
equations for $f_{\nu_{\alpha}}$ and $f_{\nu_s}$ that reproduce
the QKE behaviour exceedingly well. This technique has been called
the ``static approximation'' \cite{ftv,fv1}. From another
perspective, the resulting equations have been shown to arise in
the adiabatic limit approximation to the QKEs in the large damping
or collision dominated epoch \cite{bvw,new1,new2}. The approximate
equations feature only the distribution functions, not the
coherences $P_{x,y}$, so we are working in a Boltzmann limit.

We will not repeat the derivations in Refs.\
\cite{ftv,fv1,bvw,new1,new2}, but present just the results.
Define:
\begin{equation}
z_{\alpha,s}\equiv \frac{f_{\nu_{\alpha,s}}}{f_{\rm eq}^{0}},\quad
 \overline{z}_{\alpha,s}\equiv
\frac{f_{\overline{\nu}_{\alpha,s}}}{f_{\rm eq}^{0}}.
\end{equation}
The $z's$ are the actual $\nu_{\alpha}$ and $\nu_s$ distribution
functions normalised by the Fermi--Dirac distribution with zero
chemical potential. In the static approximation, the sterile
neutrino $z$'s obey the rate equations
\begin{equation}\label{eq:zs}
\frac{\partial z_{s}}{\partial T}= -\frac{\Gamma_{\alpha s}}{{\cal
H}T}\left[z_{\alpha}-z_{s}\right], \quad \frac{\partial
\overline{z}_{s}}{\partial T}= -\frac{\overline{\Gamma}_{\alpha
s}}{{\cal H}T} \left[\overline{z}_{\alpha}-\overline{z}_{s}\right]
\label{zeq},
\end{equation}
where
\begin{equation}\label{eq:gamma}
\Gamma_{\alpha s} = \frac{1}{4}\frac{\Gamma s^{2}}
{s^{2}+d^{2}+(c-b+a)^{2}},\quad \overline{\Gamma}_{\alpha
s}=\frac{1}{4}\,\frac{\Gamma s^{2}} {s^{2}+d^{2}+(c-b-a)^{2}},
\end{equation}
are the transition rates.

Using lepton number conservation and the following definitions:
\begin{equation}
z^{+}_{s}\equiv\frac{z_{s}+\overline{z}_{s}}{2},\quad
z^{-}_{s}\equiv\frac{z_{s}-\overline{z}_{s}}{2},
\end{equation}
we see that the neutrino asymmetry evolution equation is given by
\begin{equation}
\label{bingo} \frac{dL_{\nu_{\alpha}}}{dt}=-\frac{1}{n_{\gamma}}
\frac{T^3}{2 \pi^2} \int^{\infty}_0  \frac{\partial
z_{s}^{-}}{\partial t}\ f^{0}_{\rm eq}\ y^2 dy .
\end{equation}
From Eq.\ (\ref{zeq}) the rate of change of $z_s^{-}$ is obtained
by solving the coupled equations
\begin{eqnarray}\label{eq:zpiu}
\frac{\partial z^{+}_{s}}{\partial t}&=&\frac{\Gamma_{\alpha s}+
\overline{\Gamma}_{\alpha s}}{2}[z^{+}_{\alpha}-z^{+}_{s}]+
\frac{\Gamma_{\alpha s}-\overline{\Gamma}_{\alpha s}}{2}
\left[z^{-}_{\alpha}-z^{-}_{s}\right], \nonumber \\ \frac{\partial
z^{-}_{s}}{\partial t}&=& \frac{\Gamma_{\alpha
s}-\overline{\Gamma}_{\alpha s}}{2}[z^{+}_{\alpha}-z^{+}_{s}]+
\frac{\Gamma_{\alpha s}+\overline{\Gamma}_{\alpha s}}{2}
\left[z^{-}_{\alpha}-z^{-}_{s}\right].
\end{eqnarray}
Adopting the instantaneous repopulation approximation
\begin{equation}
z_{\alpha} =  \frac{ 1 + e^y }{ 1 + e^{y - \tilde{\mu}_{\alpha}}
}, \quad \overline{z}_{\alpha} =  \frac{ 1 + e^y }{ 1 + e^{y +
\tilde{\mu}_{\alpha}} },
\end{equation}
and expanding these expressions to first order in the chemical
potential, we finally obtain
\begin{equation}\label{eq:ABC}
\frac{dL}{d(T/{\rm MeV})}=\int^{\infty}_0
{AL+BL_{\nu_{\alpha}}+B'\over \Delta(s^2)} \ dy,
\end{equation}
with
\begin{eqnarray}\label{eq:A}
A &\simeq &  -1.27 s^{2} k_{\alpha}\left({{\rm eV}^2\over \delta
m^2}\right)
                        \left(\frac{T}{\rm MeV}\right)^6
                (b-c) y^{4} (1+e^{y})^{-1}(1-z^{+}_{s}),
                \nonumber
\\
 B & \simeq &  0.09 s^{2} k_{\alpha} \left(\frac{T}{\rm MeV}\right)^2
               [s^{2}+d^{2}+(c-b)^{2}+a^{2}]
            y^{3} e^{y} (1+e^{y})^{-2}, \nonumber
\\
B'  & \simeq &  -0.03 s^{2} k_{\alpha} \left(\frac{T}{\rm
MeV}\right)^2  [s^{2}+d^{2}+(c-b)^{2}+a^{2}] y^{3}
        (1+e^{y})^{-1}z^{-}_{s},
\end{eqnarray}
where we have used $n_{\gamma} = 2 \zeta (3) T^3/\pi^2 \simeq
T^3/4.1$, and
\begin{equation}
\Delta(s^2)=[s^{2}+d^{2}+(c-b+a)^{2}]\,[s^{2}+d^{2}+(c-b-a)^{2}].
\end{equation}
Equations (\ref{eq:zpiu}) and (\ref{eq:ABC}) form a coupled system
that must be simultaneously solved to track asymmetry evolution
and sterile neutrino production.

\subsection{Repopulation, the damped MSW effect,
and the adiabatic limit in the collision dominated epoch}
\label{sec3B}

References \cite{bvw,new1,new2} show how the static approximation
follows from the QKEs in the adiabatic limit. These results are
briefly reviewed, with the focus on repopulation handling. We will
see that repopulation is not a critical issue for asymmetry
evolution in the collision dominated regime, although it does play
an interesting role in the otherwise mundane $T>T_c$ epoch.
Importantly, we explain how the static approximation excludes the
MSW effect.

One may extract the collision dominated adiabatic limit beginning
with the four coupled QKEs [Eq.\ (\ref{eq:b1})] \cite{new2}. These
calculations show that the proper incorporation of repopulation
affects the evolution of $L_{\nu_{\alpha}}$ at sub-leading order
only; an excellent description of the asymmetry evolution in the
said limit may still be obtained from less rigorous repopulation
treatments. We will not repeat here the somewhat involved algebra
of Ref.\cite{new2},  but instead explain the gist of the result
through a simpler argument. The key is to consider repopulation in
two extreme limits: a {\it full decoupling} limit and a {\it
thermal equilibrium} limit.

The first limit corresponds to completely switching off the
refilling interactions, i.e., $\frac{\partial P_{0}}{\partial
t}=0$ \cite{bvw}, while the other assumes $f_{\nu_{\alpha}}=f_{\rm
eq}$, where the right hand side includes the appropriate chemical
potential \cite{ropa}. In comparison, the latter limit is much
closer to the truth since weak interaction induced scattering
occurs frequently in the epoch under consideration.

In the full decoupling limit, the QKEs [Eq.\ (\ref{eq:b1})] reduce
to the homogeneous system \cite{bvw}
\begin{equation}
\label{decoupled}
{\partial {\bf P}\over \partial t}={\cal K}{\bf
P},
\end{equation}
where the $3\times 3$ matrix ${\cal K}$ is given by
\begin{equation}
{\cal K}=\left(
\begin{array}{ccc}
-D   & -V_z & 0    \\
 V_z & -D   & -V_x \\
 0   & V_x  & 0
\end{array} \right)
={1 \over \ell_0}\,
\left(
\begin{array}{ccc}
-d   & (c\pm a-b) & 0    \\
 -(c\pm a-b) & -d   & -s \\
 0   &  s  & 0
\end{array} \right).
\end{equation}
Rewriting Eq.\ (\ref{decoupled}) in the instantaneous diagonal
basis, and taking the adiabatic limit by ignoring the time
derivatives of the diagonalisation matrix \cite{bvw,new1,new2},
the evolution is now driven by the eigenvalues $\lambda_{1,2,3}$
of ${\cal K}$. In most cases, $\lambda_{1,2}$ form a complex
conjugate pair. Since we are interested in the collision dominated
regime which includes requiring that $V_x/D \ll 1$, these are well
approximated by
\begin{equation}
\lambda_{1,2} \simeq - D \pm i \sqrt{V_x^2 + V_z^2} = - \ell_{\rm
int}^{-1} \pm i \ell_{m}^{-1}.
\end{equation}
The remaining eigenvalue $\lambda_3$ is small, real and negative,
and obeys \cite{new1}
\begin{equation}
\lambda_3 = - \frac{D V_x^2}{(D-\lambda_3)^2 + V_x^2 +
V_z^2}\simeq - \frac{D\,V^2_x}{D^2 + V_x^2 + V_z^2} = - D \frac{
\left(\frac{V_x}{D}\right)^2 } { 1 + \frac{ V_x^2 + V_z^2 }{ D^2 }
}.
\end{equation}
Since the real parts of $\lambda_{1,2}$ are negative and much
larger in magnitude than $\lambda_3$, the solution lies in the
direction of the eigenvector associated with $\lambda_3$, while
the transverse components are strongly damped (which is
 another way to define collision dominance
\cite{bvw,new1}).

It is very important to observe that the oscillatory aspect of the
evolution, driven by the imaginary parts of $\lambda_{1,2}$, is
severely damped in the regime concerned. {\it This is precisely
where the pure MSW effect is removed from the dynamics. The DHPS
procedure also contains a step which nullifies MSW transitions, a
point they failed to notice.} Note that the ratio of the imaginary
to the real parts of $\lambda_{1,2}$ is approximately $\ell_{\rm
int}/\ell_m$. When this ratio is not small, many oscillations
happen within one interaction length, thus allowing coherent MSW
transitions to take place, particularly at low temperatures where
$\ell_{\rm int}$ is large. Using the diagonalisation matrix, one can
also see that the oscillation amplitude is large.
See Refs.\ \cite{bvw,new1} for further
discussions on extracting the MSW effect from the QKEs.

The sterile neutrino production equation obtained in the fully
decoupled limit is simply
\begin{equation}{\label{eq:zslamb}}
{\partial z_{s}\over \partial t} \simeq -{1\over 2} \lambda_3
[z_{\alpha}-z_s],
\end{equation}
which is in agreement with the static approximation. The related
asymmetry evolution equation is also of static approximation form.

In the opposite thermal equilibrium limit, the QKEs become also a
homogeneous system of three  differential equations if chemical
potentials are ignored in the active neutrino distribution
functions, so that $\frac{\partial z_{\alpha}}{\partial t}=0$ and
thus $\frac{\partial P_{0}}{\partial t}=-\frac{\partial
P_{z}}{\partial t}$, with the matrix ${\cal K}$ given by
\begin{equation}
{\cal K}=
\left(
\begin{array}{ccc}
-D   & -V_z & 0    \\
 V_z & -D   & -V_x \\
 0   & \frac{V_x}{2}  & 0
\end{array} \right)
={1 \over \ell_0}\,
\left(
\begin{array}{ccc}
-d   & (c\pm a-b) & 0    \\
 -(c\pm a-b) & -d   & -s \\
 0   &  \frac{s}{2}  & 0
\end{array} \right).
\end{equation}

Repeating the same procedure as before yields only one difference:
the denominator of $\lambda_3$ now requires the substitution
$V_x^2\rightarrow V_x^2/2$. This, however, has no practical
importance in the collision dominated regime, since the condition
$\left. r \right|_{\rm res} \equiv \left. ({\ell}_{\rm
int}/{\ell}_{m})\right|_{\rm res} = \left. (V_x/D) \right|_{\rm
res}\ll 1$ is required to validate the approximate equations at
resonance. As $\left. r \right|_{\rm res} \to 1$, collision
dominance collapses locally; MSW transitions become the chief
amplification mode \cite{fv1,qke,bvw,new1,new2}.

To conclude, we find that the rate of sterile neutrino production
and asymmetry generation is the same at leading order in $V_x^2$,
independent of the description of the active neutrino
distributions: full decoupling \cite{bvw}, thermal equilibrium
\cite{ropa}, or the more general equation \cite{new2}. Deviations
from the initial thermal equilibrium distribution have no
noticeable effects on the asymmetry generation mechanism. This
statement is qualitatively understood, because strongly damped
oscillations do not produce significant spectral distortions.

\subsection{Comparison of the QKEs and the static approximation}
\label{sec3c}

The dash-dotted curves in Figs.\ 1a and 2a show the integration of
the static approximation equations (\ref{eq:zpiu}) and
(\ref{eq:ABC}), which are in excellent agreement with the QKE
results (solid lines) prior to and during explosive growth. The
essential physics in the collision dominated regime is very well
captured by the static approximation: the MSW effect is damped,
and  rate equations for the neutrino distribution functions govern
the evolution. In terms of the QKEs' adiabatic limit, the
transition rates in these equations are simply the eigenvalue
$\lambda_3$.

Moreover, one can easily infer from Eq.\ (\ref{eq:ABC}) the
existence of a critical temperature $T_c$ and its approximate
value by examining the
 dominant term $A$
\cite{ftv,fv1,bvw}. This term has a fixed point at $L = 0$ that is
stable above $T_c$, and unstable below it where deviations from $L
= 0$ will cause runaway positive feedback. The transition is
controlled by the function $(b - c)$ in Eq.\ (\ref{eq:A}). Since
$b \sim T^6$, the $A$ term is generally positive for $\delta m^2 <
0$ at high temperatures, and $L$ is destroyed. At $T_c$, the
``effective $A$'' [the $y$ integration in Eq.\ (\ref{eq:ABC}) must
be performed to determine the net effect of $A$] becomes negative,
leading to a spurt of exponential growth.

The subdominant $B$ and $B'$ terms in Eq.\ (\ref{eq:ABC}) control
the extent to which the condition $L=0$ is approximated prior to
$T_c$; realistically $L = 0$ does not exactly define a fixed point
since $B$ and $B'$ are not proportional to $L$. Numerically, we
find that $L \sim 10^{-15}$ is a typical value. Interestingly,
DHPS report values as low as $10^{-100}$. The reason for this is
clear: the $B$ and $B'$ terms in Eq.\ (\ref{eq:ABC}) arise from
the nonzero chemical potentials in $z_{\alpha}$ and
$\overline{z}_{\alpha}$. These terms are absent in the DHPS
equations since chemical potentials are neglected in the
repopulation term, even when they attempt to numerically solve the
QKEs [see Eq.\ (9) of DHPS and following discussion].  In their
case $L = 0$ is much closer to being a true fixed point. This is a
relatively minor issue, and not germane to the controversy over
the final asymmetry value. Nevertheless, it is of some numerical
interest since tracking values of order $10^{-15}$, rather than
$10^{-100}$, is a much easier task. Also, if $10^{-100}$ was the
correct number, one must consider the presence of statistical
fluctuations in the early universe as they would be able to change
randomly the sign of the asymmetry at different spatial points
\cite{ftv,dolgov,ropa}. In the presence of chemical potentials in
the active neutrino distributions, however, this effect can be
excluded \cite{ropa} for a large range of mixing parameters
(roughly for $\sin^2 2\theta_0\lesssim 10^{-6}$).

On the practical front, the neglect of chemical potentials when
numerically integrating the QKEs leads to an artificially higher
sensitivity to small effects on the solution (numerical error,
oscillatory terms). DHPS observed this sensitivity, but did not
realise that the inclusion of chemical potentials was remedial
from this perspective. This may be one reason why DHPS found
that their results were
chaotic in sign for some mixing parameters, while an analysis
performed  with chemical potentials did not
\cite{ropa}.
DHPS took this synthetic chaoticity as grounds to not
rely on their own numerical solutions to the QKEs, or those of others
\cite{qke,ropa}.\footnote{Curiously, DHPS relied on the results of
Ref.\ \cite{kirilova}, which also reported rapid oscillations 
in the sign of the neutrino
asymmetry for $\delta m^2< 10^{-7}$ eV$^2$, although the
equations used in Ref.\ \cite{kirilova} are no more accurate
 than those in Refs.\ \cite{qke,ropa}. DHPS may not have been aware
of this incongruity since  Ref.\ \cite{kirilova} does not contain
any figure of the asymmetry behaviour, although rapid sign changes
are clearly reported.}

After explosive growth, Figs.\ 1a and 2a show that the static
approximation equations significantly underestimate the subsequent
augmentation of $L$.  This is the epoch in which the adiabatic MSW
effect takes over as the principal propellant.

\subsection{MSW dominated regime}

Collision dominance and thus the static approximation collapse
when ${\ell_{\rm int}}/{\ell_{m}}$ approaches and becomes larger
than one \cite{fv1,qke}. This is particularly evident in the
adiabatic limit approach \cite{bvw,new1}.  Consider the
eigenvalues of the matrix ${\cal K}$ at resonance:
\begin{equation}
\left. \lambda\right|_{\rm res} = -D,\qquad
- \frac{D}{2} \pm i \frac{\sqrt{4 V_x^2 - D^2}}{2}.
\end{equation}
When collision dominance breaks down, i.e., $\left.
(V_x/D)\right|_{\rm res} \gtrsim 1$, the oscillatory imaginary
parts are {\it not} preferentially damped. The real eigenvalue
$\lambda_3$ and the real parts of the complex conjugate pair are
now comparable in size.  The imaginary components drive the MSW
effect \cite{bvw,new1}, inducing many large amplitude
oscillation cycles within the decoherence time scale.

The simplest way to explore the implications of the MSW effect is
through the approach of Ref.\cite{qke}. We begin with a very
important observation: after the initial burst of $L$ creation,
the neutrino and antineutrino resonances become separated in
momentum space. This is illustrated in Figs.\ 1b and 2b, and is
easy to see analytically from Eqs.\ (\ref{cequalsb}) and
(\ref{cequalspma}). At $T_c$, Eq.\ (\ref{cequalsb}) holds, with
neutrinos and antineutrinos having the same resonance momentum.
After a substantial $L$ has been created, the resonance condition
evolves to
 being given by Eq.\ (\ref{cequalspma}),
and the resonance momenta are separated. If $L > 0$, then the
neutrino resonance quickly moves to the tail of the distribution,
while the antineutrino resonance stays in the body. The MSW effect
therefore has very asymmetric consequences for neutrinos and
antineutrinos: the former are unaffected, while the latter
continue to be processed into sterile states, thereby adding to
the asymmetry. We shall henceforth assume $L > 0$ for
definiteness.

Let us focus now on the large region of parameter space for which
the MSW transitions are adiabatic. Here, complete conversion of
antineutrinos at the resonance momentum into sterile states means
that the asymmetry is augmented, while the position of the
resonance evolves with $L$. Indeed, the rate of asymmetry growth
depends on how quickly the resonance moves through the
distribution. In the case of a small resonance width, it is easy
to make this connection heuristically, which in turn gives rise to
a simple evolution equation \cite{qke}
\begin{equation}
\frac{ d L_{\nu_{\alpha}} }{ dT } = - \frac{1}{n_{\gamma}}
\frac{T^3}{2\pi^2}\, [f_{\overline{\nu}_{\alpha}}(y_{\rm res}) -
f_{\overline{\nu}_{s}}(y_{\rm res})]\, y^2_{\rm res} \frac{ d
y_{\rm res} }{ dT } \equiv - X  \frac{ d y_{\rm res} }{ dT }.
\label{preseq}
\end{equation}
Using the resonance condition of Eq.\ (\ref{cequalspma}), one
obtains from Eq.\ (\ref{preseq}) the expression \cite{qke}
\begin{equation}
\frac{ d L_{\nu_{\alpha}} }{ dT } = - \frac{4}{T} \frac{X y_{\rm
res}}{ 1 + \frac{X y_{\rm res}}{L_{\nu_{\alpha}}} }, \label{LMSW}
\end{equation}
as a non-linear asymmetry evolution equation, for the case where
$\frac{d y_{\rm res}}{d T} < 0$. Repopulation is incorporated as
per the procedure described in Ref.\cite{bfv}.

The result of solving Eq.\ (\ref{LMSW}) was first presented in
Fig.\ 1 of Ref.\ \cite{qke}, and compared with the QKE solution
for temperatures $T \lesssim T_c/2$. Excellent agreement was
obtained, and the final asymmetry was found to be in the $0.2 -
0.37$ range depending on the oscillation parameters. {\it
Numerical solutions of the QKEs and  of Eq.\ (\ref{LMSW}) yielded
completely consistent results for the temperature range where the
latter equation was expected to be a good approximation.}

Notice in particular that when $L_{\nu_{\alpha}} \ll 1$, Eq.\ (\ref{LMSW})
simplifies to
\begin{equation}
\frac{ d L_{\nu_{\alpha}} }{ dT } \simeq - \frac{4
L_{\nu_{\alpha}}}{T},
\end{equation}
which immediately implies that $L_{\nu_{\alpha}} \propto T^{-4}$
in that regime. {\it The $T^{-4}$ behaviour that can be
empirically measured from the QKE curves is fully explained by the
adiabatic MSW effect.}

As the asymmetry continues  to grow, the antineutrino resonance
evolves out of the body into the high momentum end of the
distribution. The asymmetry becomes frozen at some final steady
state value, whose approximate magnitude can be easily understood
by integrating the Fermi--Dirac distribution from $y = y_{\rm
res}^{\rm low} \sim 0$ to $y = \infty$:
\begin{equation}
\label{freeze}
L^{\rm final}_{\nu_{\alpha}} \sim
\frac{1}{4\zeta(3)}
\left(\frac{T_{\nu_{\alpha}}}{T_{\gamma}}\right)^3 \int_0^{\infty}
\frac{ y^2 dy}{1 + e^y}
 = \frac{3}{8}\left(\frac{T_{\nu_{\alpha}}}{T_{\gamma}}\right)^3.
\end{equation}
The temperature ratio takes care of reheating due to $e^{+}e^{-}$
annihilations at $T \simeq m_e \simeq 0.5$ MeV. Equation
(\ref{freeze}) gives the approximate magnitude only, because the
distribution changes with time as the asymmetry is created.
Numerically, the final values found by incorporating proper
thermalisation effects are quite close to this estimate.

\subsection{Combining the collision dominated and MSW dominated regimes.}

Figures 1a and 2a summarise neatly the overall picture of
asymmetry evolution, as discussed in Sec.\ \ref{sec1}.
 The solid and dash-dotted lines represent solutions to the exact QKEs and
 the static approximation equations respectively. Clearly, the latter
 offers an excellent description for
the explosive growth phase, but underrates the subsequent growth
of $L_{\nu_{\alpha}}$. The dashed line is generated by a code that
computes the ratio between the resonance interaction and
matter-affected oscillation lengths, $r|_{\rm res} \equiv \left.
({\ell_{\rm int}}/{\ell_m}) \right|_{\rm res}$, at every time
step. If $r|_{\rm res} <1$, the static approximation equations are
used to advance the system. For $r|_{\rm res}
> 1$,  pure adiabatic MSW evolution is employed
via Eq.\ (\ref{LMSW}). In both the high and low $T$ regimes, the
dashed line and the QKEs are in superb agreement. The small
disparity directly after  explosive growth arises from the fact
that neither effect is dominant in this intermediate stage. We
conclude that the physics of asymmetry growth has been thoroughly
understood.

\section{Comparison with the DHPS results}
\label{sec4}

 In Sec.\ \ref{sec4A} we ``deconstruct'' the DHPS
equations, while we devote Sec.\ \ref{fatal} to pinpointing where
in their derivation the MSW effect is unknowingly neglected.

\subsection{An analysis of the DHPS equations}
\label{sec4A}

For comparison with the static approximation, we rewrite DHPS's
evolution equations in the notation of this Comment using the
``conversion relations'' given in the Appendix.

Equation (42) of DHPS describes sterile neutrino production. Its
translation is
\begin{equation}
{\partial z^+_s\over \partial (T/{\rm MeV})}=-{s^2\Gamma\over
4{\cal H}T} {d^2+(b-c)^2+a^2 \over \Delta(0)}[1-z^+_s].
\label{DHPS42}
\end{equation}
Comparing this with Eq.\ (\ref{eq:zpiu}), one sees that it agrees
with the latter's leading term, when (i) chemical potentials are
neglected and thus $z_{\alpha}^+ \simeq 1$, and (ii) the
replacement
\begin{equation}
\Delta(s^2)\rightarrow \Delta(0), \label{substitution}
\end{equation}
is made.

The neutrino asymmetry evolution equation [Eq.\ (51) of DHPS] is
equivalently
\begin{equation}
{1\over L}\,\frac{dL}{d(T/{\rm MeV})}=\int_0^{\infty}
{A\over\Delta(0)} \left[1-{s^2\over 4}
\frac{d^2+(c-b)^2+a^2}{\Delta(0)} \right] dy. \label{DHPS51}
\end{equation}
The first term on the right hand side is again connected to the
leading $A$ term of the static approximation [Eq.\
(\ref{eq:ABC})], with the substitution Eq.\ (\ref{substitution}).
The $B$ and $B'$ terms are absent as expected, since DHPS neglect
chemical potentials in the $\nu_{\alpha}$ and
$\overline{\nu}_{\alpha}$ distribution functions.

The second term in the square brackets is DHPS's back-reaction
term $B_1$.  This supposedly new term has, in fact, an interesting
interpretation. Consider the function $\Delta(s^2/8)$. One can
perform the expansion
\begin{eqnarray}
{A\over \Delta(s^2/8)} &=&  {A\over \Delta(0)} \left[1-{s^2\over
4}\frac{d^2+(c-b)^2+a^2}{\Delta(0)}\right] \nonumber \\
 & & \quad+ {\cal O}\left({s^2\over d^2+(c-b+a)^2},{s^2\over d^2+(c-b-a)^2}\right)^2.
\label{expansion}
\end{eqnarray}
Thus the back-reaction term is but the first order correction in
an expansion of $\Delta(s^2/8)$. But we have already discussed in
Sec.\ \ref{sec3B} this sort of expansion,  in the context of the
$V_x^2$ term in the denominator of the eigenvalue $\lambda_3$,
where the {\it full decoupling} and the {\it instantaneous
repopulation} limits yielded $\Delta(s^2)$ and $\Delta(s^2/2)$
respectively. We argued that this distinction was of no practical
importance, because the numerical difference between these two
cases was only significant when the collision dominance assumption
used in their derivation broke down. In a nutshell: (i) The first
order terms are only large when the MSW effect is important. (ii)
When the MSW effect is important, equations such as obtained from
the static approximation, and DHPS (51), are no longer a good
approximation to the QKEs.

Let us now evaluate Eq.\ (\ref{expansion}) for the antineutrino
($c - b - a = 0$) resonance:
\begin{equation}
\left.{A\over \Delta(s^2/8)}\right|_{\rm res} \simeq
\left. {A\over \Delta(0)}\right|_{\rm res}
\,\left[1-{s^2\over 4}\frac{d^2_{\rm res}+2a^2}
{d^2_{\rm res}(d^2_{\rm res}+4a^2)}\right].
\end{equation}
When the asymmetry grows at the critical temperature $T_c$, the
resonance condition passes from a regime in which $b \simeq c$
when $L$ is negligible, to $a \simeq c$. In this second regime the
term $d_{\rm res}^2\ll a^2$ and  the last expression becomes
simply,
\begin{equation}
\left.{A\over \Delta(s^2/8)}\right|_{\rm res} \simeq \left.
{A\over \Delta(0)}\right|_{\rm res} \left[1-{s^2\over 8 d^2_{\rm
res}}\right] .
\end{equation}
So we see that the expansion parameter on resonance is $s^2/d_{\rm
res}^2$, i.e., the square of the ratio between the local
interaction and matter-affected oscillation lengths $(\left.
r\right|_{\rm res})^2$. As explained earlier, a small $\left.
r\right|_{\rm res}$ denotes that collisions are frequent enough to
interrupt coherent evolution and thus to prevent the MSW effect,
while a large $\left. r\right|_{\rm res}$ means MSW conversion can
occur as the collisional mechanism switches off. Of course this
transition is gradual.

Moreover, using the equality $d=\delta b$, one can also see, using
Eqs.\ (\ref{eq:b}) and (\ref{eq:yres}), that
\begin{equation}
\label{whatever}
\frac{s^2}{d_{\rm
res}^2}=\left({s\over\delta}\right)^2 \left({y_{\rm res}^0\over
y_{\rm res}}\right)^4= \left({s\over\delta}\right)^2
\left({|L|\over L_t} \right)^3,
\end{equation}
where $y_{\rm res}^0$ is the resonant momentum when the asymmetry
is negligible. Recall that $y_{\rm res}^0$ is very small at high
$T$, but grows as the temperature decreases, reaching $\sim 2$ at
$T_c$. The quantity $y_{\rm res}$ is the general resonance
momentum, which attains a minimum value depending on $\delta m^2$
(smaller for low $\delta m^2$) when the asymmetry grows
 (see Figs.\ 1b and 2b). It is thus clear from Eq.\ (\ref{whatever})
  that if one continues to use $B_1$ in the
evolution equation when the asymmetry is large and $s^2/d_{\rm
res}^2 \sim {\cal O}(1)$, the rise of the asymmetry will be
artificially made slower. From that moment its evolution will be
approximately $|L|\sim 50\ L_t$ and
from the equations Eq.(\ref{eq:Lt}) and Eq.(\ref{eq:yres}) this implies
that $|L|\propto 1/T$,
a behaviour observed by DHPS in their results. In
this way the solution can grow just one order of magnitude since
the onset of the instability.

In summary, the back-reaction term $B_1$ is generated by an
expansion in $s/d_{\rm res}$,  and not simply $s/\delta$ as DHPS
believe. {\it Their evolution equation is therefore invalid for
temperatures at which $s^2/d_{\rm res}^2> 1$.}
As emphasised before \cite{fv1,bvw}, both the
static approximation and the DHPS equations belong to a family of
collision dominated idealisations which require $s/d_{\rm} \ll 1$
(plus other conditions) for their validity. When $s/d \gtrsim 1$,
the MSW effect is missed in both cases. Clearly DHPS did not
properly appreciate the constraints on the real expansion
parameter adopted in their derivation, and hence trusted their
equations' validity for all temperatures.

For very large $\delta m^2$, DHPS do actually observe a large final
value. This is because $|L|_{\rm fin}\simeq
10^{-5} (|\delta m^2|/{\rm eV}^2)^{1/3} (T_c/T_f) \simeq
10^{-4} (|\delta m^2|/{\rm eV}^2)^{1/3}$ where $T_f$ is the freezing
temperature corresponding to a resonant momentum
(of antineutrinos if $L$ is positive, of neutrinos if $L$ is
negative) well in the tail of the distribution ($y_{\rm res}^f
\gtrsim
10$) and is very weakly dependent on the exact choice of $y_{\rm res}^f$.
Thus for
$|\delta m^2|\simeq 10^9\ {\rm eV}^2$ they find $|L|_{\rm fin}\sim 0.1$,
and are therefore still able to state that active--sterile neutrino
oscillations can yield large asymmetries.
{\it However, this high $\delta m^2$ asymmetry rise is spurious. For
$\delta m^2\gtrsim 10^4\ {\rm eV}^2$,
 the effective potential as given by Eq.\ (\ref{eq:effp})
is meaningless, since the critical temperature is now much larger
than $100$ MeV. The conditions of the plasma are different,
especially at temperatures above the quark--hadron phase
transition. The dynamics of neutrino oscillations in this regime
is actually an open problem.}

\subsection{How DHPS neglect the MSW effect}
\label{fatal}

 In extracting their simplified evolution
equations from the QKEs, DHPS make use of two small expansion
parameters: (i)  $1/ Q$, and (ii)
 $\sin 2 \theta$.  We now examine their merits.

\medskip
\noindent 1. {\it Small $1/Q$ expansion}. The derivation begins
with an attempt to solve analytically two out of a system of eight
 coupled differential equations, namely, Eqs.\ (27) and (28) in
DHPS:
\begin{eqnarray}
\label{LHDEs} h_-'/Q &=& U l_- - \gamma h_- - V Z l_+, \nonumber
\\  l_-'/Q & = & \frac{F}{2}(a_- - s_-)
- U h_- - \gamma l_- + V Z h_+,
\end{eqnarray}
where prime denotes differentiation with respect to $\tau$. These
expressions are equivalent to
\begin{eqnarray}
\frac{2p}{\delta m^2} \frac{\partial}{\partial t} (P_x
-\overline{P}_x) &=& (1 - b) (P_y - \overline{P}_y) -d (P_x -
\overline{P}_x) +  a (P_y + \overline{P}_y), \nonumber \\
\frac{2p}{\delta m^2} \frac{\partial}{\partial t} (P_y
-\overline{P}_y) &=&-s (P_z - \overline{P}_z) -( 1 - b) (P_x -
\overline{P}_x)-d (P_y - \overline{P}_y)-a (P_x + \overline{P}_x),
\end{eqnarray}
in our language, with $c \simeq 1$. Solutions to Eq.\
(\ref{LHDEs}) are generically oscillatory.  To this end, DHPS
devise an oscillation averaging procedure that consists of setting
all terms containing $1/Q$ (i.e., the $\tau$-derivatives) to zero,
from which they obtain what are actually ``steady state''
solutions to the variables $h_-$ and $l_-$.  We now explain why
this method is flawed.

Firstly, the variable $P_y$ enters the big picture as the
``driving force'' for $\nu_s$ production $\frac{\partial
z_s}{\partial t}$ [see Eqs.\ (\ref{formaldLdt}) and
(\ref{bingo})]. In the context of solving differential equations,
one may ignore contributions from oscillatory components in the
driving force $P_y$ {\it only} if the oscillation amplitude is
much smaller than the size of its steady ``main term''. Otherwise,
oscillations propel the evolution, producing a marked change in
the {\it integrated} variable $z_s$ (i.e., the MSW effect) when
the oscillation frequency is a minimum at resonance, in the same
way that $\frac{d}{dt} \left(\frac{1}{\omega} \sin \omega t
\right) = \cos \omega t$.

In their derivation, DHPS do not consider at all the magnitudes of
these oscillatory terms, and discard them on the basis that they
impair computational efficiency, when in fact even the adiabatic
solution to $P_y$ (and $\overline{P}_y$) is purely oscillatory at
resonance in the post explosive growth environment where a
relatively large $r|_{\rm res} \equiv (\ell_{\rm
int}/\ell_{m})|_{\rm res}$ ratio prevails \cite{bvw,new1}. {\it
Their equations cannot describe the sharp change in $z_s$ at the
low temperature resonance because the dominant and strongly
frequency-dependent term responsible for the MSW effect has been
removed.}  Indeed, one cannot know the sizes of the steady and
oscillatory components in $P_y$ or $l_-$ without some input of the
initial conditions; to eliminate dynamical variables in the
equation's differential form is premature and may be
dangerous.\footnote{We also mention a point noted in Ref.\
\cite{sorri}, that when the asymmetry has grown sufficiently to
separate the neutrino and antineutrino resonances, the differences
in the distributions, and their derivatives, can be as large as
their sums, simply because one distribution is on resonance while
the other is far from it.}

Moreover, the process of artificially factoring out a large
quantity by redefining the integration variable and then
exploiting its inverse as an expansion parameter is not in fact a
well-justified man{\oe}uvre to obtaining steady state solutions.
To see this, one needs only to consider the case of exponential
growth/decay, $\frac{d x}{d t} = A x$, where the approximation
$\frac{1}{A}\frac{dx}{d t} \to 0$ and thus the conclusion $x
\simeq 0$  crumble for a positive $A$.

Fortunately, steady state solutions to $h_-$ and $l_-$ are indeed
valid for some oscillation parameters $(\delta m^2, \sin 2
\theta)$ when collisions are frequent enough to quickly damp the
oscillations, even at resonance, provided that the
 adiabatic condition is also
satisfied \cite{bvw,new1}. Thus DHPS have managed to reproduce
results (albeit from a somewhat dubious procedure) that are
consistent with the static approximation, in the manner that the
latter may also be derived in the $d \gg s$ limit by setting
$\frac{d P_x}{dt} \simeq \frac{d P_y}{dt} \simeq 0$. But, like the
static approximation, their equations do not apply to the low
temperature region after explosive growth where steady state
solutions do not hold at resonance.

\medskip
\noindent 2. {\it Small $\sin 2 \theta$ expansion}. This
approximation is used on two occasions, first as a pretext for
disregarding the variable $P_z - \overline{P}_z$ (or $a_- - s_-$)
in Eq.\ (\ref{LHDEs}) [Eq.\ (28) in DHPS], and then in evaluating
the eigenvalues and other properties of the matrix ${\cal M}$ in
their Eq.\ (33).

The first exploit is precarious from the perspective of solving
differential equations, since $P_{x,y,z,0}$ (and therefore
$a_{\pm},s_{\pm},h_{\pm},l_{\pm}$) are {\it interdependent
dynamical variables} and one has no prior knowledge on their
sizes.  Indeed, it is far more seemly to assume {\it a priori} the
$P_z - \overline{P}_z$ term to be of some substance at resonance
when the MSW effect kicks in at low temperatures, given that only
antineutrinos (for $L>0$) can be converted into sterile states in
this epoch by the said mechanism. The neglect of this term is also
related to DHPS's observation of the asymmetry reaching
arbitrarily small values prior to explosive growth, to be
discussed later in Sec.\ \ref{sec5}.

In the second instance, the expansion procedure employed to
diagonalise the matrix
\begin{equation}
\label{matrix} {\cal M} = \left( \begin{array}{cccc}
                0 & 0 & 0 & F \\
                0 & -2 \gamma & 0 & -F \\
                0 & 0 & -\tilde{\gamma} & \tilde{U} \\
                -\frac{F}{2} & \frac{F}{2} & - \tilde{U} & - \tilde{\gamma}
                \end{array} \right),\; \tilde{U} = U\left(1 -
                \frac{D^2}{\gamma^2 + U^2} \right),\,
                \tilde{\gamma} = \gamma \left(1 +
                \frac{D^2}{\gamma^2 + U^2} \right),
\end{equation}
where $F \doteq -s$, $\gamma = \delta/\tau^2 \doteq d$, $D \doteq
-a \propto L$ and $U = 1/\tau^2 -1 \doteq 1-b$, is not itself at
flaw. However, it does have its own limitations which DHPS did not
notice.

Equation (\ref{matrix}) suggests that the term $F$ (or $s$) should
be compared with $\gamma$, $U$, $\tilde{\gamma}$ and $\tilde{U}$
if it is to be treated as a perturbation. DHPS consider only $s$
and $\delta$, presumably because $\gamma = \delta$ at resonance,
which they define as occurring at $\tau =1$ such that $U=0$ (or $c
=b$). However, this resonance condition assumes a small asymmetry,
i.e., $|b| \gg |a|$. {\it Thus it is able to describe the high
temperature regime where the term $b$ dominates, but not after a
sizeable asymmetry has been created, where the now separated
neutrino and antineutrino resonances occur at any two $\tau$'s
other than $\tau = 1$. The ratio $s/\delta$ is meaningless in this
second regime.}

Lastly, it must be stressed that MSW flavour conversion is {\it
not} a perturbative effect. {\it Small $\sin 2 \theta$ expansion
equals no MSW effect}, in the spirit that treating the
off-diagonal $\sin 2 \theta$ in the Schr\"{o}dinger Equation for a
two-flavour collisionless system as a perturbation will not yield
the correct behaviour at resonance.

\section{Other comments}
\label{sec5}

\subsection{The mystery of the oscillation in sign}

DHPS allege that results presented by other independent groups are
in conflict on the issue of oscillations in the sign of the
asymmetry at the critical temperature. They use this to cast doubt
on the credibility of some of these works. The true status of this
subject is actually very clear, as we will explain below.

To begin with, one must realise that two different systems are
discussed in the literature. The first is a simplified scenario in
which all neutrinos have the same momentum, usually taken to be
the zero chemical potential thermal average, $p = \langle p
\rangle \simeq 3.15T$. The second is the realistic case where the
neutrinos are described by a Fermi--Dirac distribution. The mean
momentum simplification should be viewed as a ``toy model'' which
is {\it not} always a good approximation to the realistic case.
However, it does give some useful qualitative insight.

Another critical issue is whether the exact QKEs  or an
idealisation such as the static approximation are used in the
calculations.

Using the mean momentum approximation, Ref.\cite{shi} observed
oscillations of sign for some parameter points and apparently drew
the incorrect conclusion that oscillations were a generic feature
of lepton number generation. The simplified system was revisited
in Ref.\cite{eks}, where oscillations were found to occur for only
part of the parameter space (see Fig.\ 2 of Ref.\ \cite{eks}). In
both papers, the mean momentum QKEs were solved\footnote{
Recently a detailed analytic study of the mean momentum model
was presented in Ref.\cite{buras}.}.

In the realistic  Fermi--Dirac case, the issue of the sign of
$L_{\nu_{\alpha}}$ was first discussed in Ref.\cite{fv1} in the
context of the static approximation;  in the region of parameter
space where this approximation is valid, the sign does not
oscillate. However, as also pointed out in that paper, the static
approximation breaks down at large $\sin^2 2\theta$ due to the
rapid rise in lepton number during the explosive growth phase.
Similar results were obtained in Ref.\cite{dibari2}.

The oscillation issue was addressed in detail in Ref.\
\cite{ropa}, where comparisons of the static approximation and the
QKEs showed that they gave concordant results for $\sin^2 2\theta
\lesssim 10^{-6}$. For larger values, sign oscillations were found
using the exact QKEs, but not the static approximation simply
because the latter is invalid for too large a $\sin^2 2 \theta$.
Also, Refs.\ \cite{ropa} and \cite{eks} are {\it not} in mutual
disagreement, since the former examined the realistic case
 with a Fermi--Dirac distribution, while the latter
considered the mean momentum approximation.

\subsection{Abrupt cut-off at large mixing angles?}

In an attempt to locate the region of parameter space for which
neutrino asymmetry generation is possible, DHPS found the
following approximate upper and lower bounds on  the so-called
``region of instability'':
\begin{eqnarray}
\sin^2 2\theta_0\cdot\sqrt{|\delta m^2|/{\rm eV}^{2}} & \lesssim &
10^{-6},  \label{eq:pippo}  \\ \sin^2 2\theta_0\cdot(|\delta
m^2|/{ \rm eV}^{2})^{1\over 6} & \gtrsim &4\times 10^{-14}.
\label{damnit}
\end{eqnarray}
While the second constraint agrees with that found in Ref.\
\cite{fv1}, the first is too severe compared with previous
calculations \cite{fv1,dibari2}.\footnote{The lower bound in Ref.\
\cite{fv1} is superficially more stringent than Eq.\
(\ref{damnit}) ($5\times 10^{-10}$ instead of $4\times 10^{-14}$)
simply because a more restrictive condition --- that the asymmetry
must grow to the $\sim 10^{-5}$ level --- was imposed in Ref.\
\cite{fv1}, while DHPS demand only amplification by as little as
one order of magnitude. However, both generate the same slope on a
$(\log \delta m^2, \log \sin 2 \theta$) graph.} In addition, DHPS
observe an abrupt transition from the ``unstable'' to the
``stable'' (i.e., no growth) regions.  This is also at odds with
earlier works, and is most likely a byproduct of numerical errors
for the following reason.

Increasing $\sin 2 \theta$ for a fixed $\delta m^2$ intensifies
sterile neutrino production, which, in turn, has the effect of
shifting the critical momentum to values larger than $2$
\cite{fv1}. By Eq.\ (\ref{eq:tres}),  the critical temperature
decreases correspondingly, and explosive asymmetry generation is
thereby delayed.  (The growth curve also becomes smoother but a
substantial asymmetry is still eventually attained.)  It  follows
that for very large mixing angles, no explosive growth can occur
since the critical momentum is then banished to the tail of the
distribution.

 DHPS note correctly  the connection between suppression of
asymmetry amplification for a large $\sin 2 \theta$ and a
substantial sterile neutrino distribution manufactured prior to
the onset of an otherwise explosive growth. The latter may be
approximated by a theta function in momentum space, such as
\cite{dibari2}
\begin{equation}\label{eq:z+}
z_s^{+} (y)= \left\{1-\exp\left[-4.3 \, (2.25)\times 10^{4} s^2
\left(\delta m^2\over {\rm eV}^2\right)^{1\over 2}\right]\right\}
\theta(y_{\rm res}^0-y),
\end{equation}
derived from Eq.\ (\ref{eq:zs}) for $\alpha=e\,(\mu,\tau)$.
Substituting Eq.(\ref{eq:pippo}) into the above, one finds that
DHPS's upper bound corresponds to having a final sterile neutrino
distribution of $z_s^+=0.04\,(0.02)$. DHPS found an expression
similar to Eq.\ (\ref{eq:z+}) in the limit $s \ll 1$ for
$\alpha=e$ [see their Eq.\ (54)].
 They argue that such accumulation of sterile neutrinos
 is able to destroy completely
the asymmetry generation since the factor $1-z_s^{+}$ appearing in
the term $A$ in the growth rate [Eq.\ \ref{eq:ABC}] now favors
stability.

As explained before, in the regime prior to the onset of asymmetry
generation, the static approximation and DHPS evolution equations
are in complete agreement at leading order; this suppression
effect is to be distinguished from DHPS's supposedly new but as
yet negligible ``back-reaction''. On the other hand, any
interested person can numerically integrate the static
approximation equations at leading order and find no abrupt
cut-off even when the sterile neutrino distribution increases to
the level of $z^+_s \simeq 0.04$.   Therefore the severity of
DHPS's upper bound has to be imputed to numerical
errors.\footnote{The calculation of $z^{+}_{s}$ should be done
numerically, coupled with the asymmetry evolution equation. The
analytical expression in Eq.\ (\ref{eq:z+}) approximates the
production with a theta function which is actually a sharp jump
around the resonance, valid only before the onset of neutrino
asymmetry generation (see Ref.\ \cite{dibari2} for more details).}

 Physically, the negative contribution
from a non-negligible sterile neutrino distribution associated
with a large $\sin 2 \theta$ is compensated for by an increase in
the critical momentum.  In more descriptive words, the asymmetry
generation ``waits'' for the onset of instability when this
compensation occurs at lower temperatures (see Ref.\
\cite{dibari2} for further discussions).

\subsection{Chemical potentials and final value of neutrino asymmetry}

As noted in the Sec.\ \ref{sec3c}, the role of chemical potentials
prior to the onset of explosive growth is to prevent the asymmetry
from being diminished to arbitrarily small values. This function
is described by the term $B$ in the static approximation [Eq.\
(\ref{eq:ABC})], or, equivalently, contained in the variable
$s(z_{\alpha}-\bar{z}_{\alpha})$ (or $Fa_{-}$ in DHPS's notation)
in the QKEs [Eq.\ (\ref{LHDEs})], which DHPS ignored.  Another
mechanism that can inhibit drastic lepton number destruction is
the presence of a sterile neutrino asymmetry, created by
oscillations from an initial $\alpha$-neutrino asymmetry. This
effect is manifested in a nonzero $s(z_{s}-\bar{z}_{s})$ (or
$Fs_{-}$)  in the QKEs, and corresponds to the term $B'$ in the
static approximation. In the case of a zero initial asymmetry,
this term is negligible. On the other hand, the term $B$ due to
finite chemical potentials must never be neglected. The fact that
DHPS omit both $B$ and $B'$ causes their solution to collapse to
values of order $10^{-100}$ prior to the onset of asymmetry
growth.

Furthermore, observe that the condition of lepton number
conservation  [Eq.\ (\ref{conservation})] and the expressions for
$P_0$ and $\overline{P}_0$ in Eq.\ (\ref{eq:b1}) together lead to
the following corollary:
\begin{equation}
\int^{\infty}_0 \Gamma(y)\left[
\left(z_{\alpha}^{\rm eq}(\mu)-
\bar{z}_{\alpha}^{\rm eq}(\bar{\mu})\right)
- \left(z_{\alpha} -
\overline{z}_{\alpha} \right) \right] f^o_{\rm eq}(y) y^2 dy
\simeq 0.
\end{equation}
Clearly, if one neglects the chemical potential $\mu_{\alpha}$,
the remaining integral is no longer vanishing, but becomes a
fictitious friction force proportional to
$-\langle\Gamma\rangle\,
L_{\nu_{\alpha}}$
in Eq.\ (\ref{formaldLdt}) through the otherwise absent
$\frac{\partial P_0}{\partial t} - \frac{\partial
\overline{P}_0}{\partial t}$ term. This point was noted in Ref.\
\cite{sorri} and speculated to be culpable for DHPS's distorted
results. However (and we are arguing in DHPS's favour), this
phantom force does not in fact appear in DHPS's final evolution
equation (51), as they specifically define  $\int^{\infty}_0
\Gamma (y) \left(z_{\alpha} - \overline{z}_{\alpha}
\right)f^o_{\rm eq}(y) y^2 dy \simeq 0$ to be their lepton charge
conservation condition [their Eq.\ (30)] which, in the absence of
chemical potentials, is actually mathematically consistent with
Eq.\ (\ref{conservation}). In effect, DHPS have taken, though
fortuitously, the
 appropriate measures to negate
 the artificial friction force  inherent
in their formulation.

\section{Conclusions}
\label{sec6}

 We review how the numerical solution to the
QKEs supplies a lepton asymmetry growth curve whose features can
be physically understood and reproduced by approximate evolution
equations. In the higher temperature collision dominated epoch,
the static approximation mimics the explosive amplification
extremely accurately, and allows one to understand why a critical
temperature exists for $\delta m^2 < 0$ and $\cos 2\theta_0 \simeq
1$. As the collision rate decreases with temperature, the MSW
effect eventually takes over as the dominant amplifier, since the
transitions are adiabatic for a large range of mixing parameters.
Also, the separation of the neutrino and antineutrino resonance
momenta after a significant $L$ has been created allows efficient
conversion of antineutrinos (for $L>0$) into sterile states, while
oscillations are matter-suppressed for neutrinos (and vice versa
for $L<0$). The roles of the static approximation and the MSW
effect are succinctly depicted in Figs.\ 1a and 2a.

The major flaw in the DHPS analysis is their neglect of the pure
MSW effect in the post explosive growth epoch. Their Eqs.\ (42)
and (51) are approximations to the QKEs valid in the collision
dominated regime and for a sufficiently small $L$. Like the static
approximation, they cannot describe the MSW effect, and therefore
underrate the asymmetry growth at lower temperatures.  However,
the DHPS underestimation is much more severe than the static case,
because their Eq.\ (51) contains a bogus back-reaction term $B_1$
that becomes large and artificially terminates the asymmetry
growth under precisely the same conditions that imply MSW
dominance. This
 can be understood from the
symbolic relation between the two approximation schemes: ${1\over
(1+x)} \simeq 1-x$, where the left hand side represents the static
situation and the right hand side correlates with DHPS. The back
reaction term $x$ leads to a severe, but fictitious, cut-off when
the MSW effect dominates at $x \sim 1$.

 We conclude that origin of the $0.2 - 0.37$ range for the final
steady state value of the asymmetry is well understood
analytically and physically, in a way which is completely
consistent with brute force numerical solutions of the Quantum
Kinetic Equations.

\acknowledgments{This work was supported by the Australian
Research Council, the Commonwealth of Australia and Istituto
Nazionale di Fisica Nucleare (INFN). We thank R. Buras, S. Hansen,
K. Kainulainen and D. Semikoz for interesting correspondence.}

\appendix

\section{Translation of DHPS notation}

We give conversion relations between the DHPS nomenclature and the
notation in this Comment, denoted by the sign ``$\doteq$'', with
DHPS on the left and our equivalent on the right.

The effective potential is
\begin{eqnarray}\label{nref}
V_{eff}^a &=& \pm C_1 \eta G_FT^3 + C_2^a \frac{G^2_F T^4
E}{\alpha} \doteq -V_{\alpha},  \nonumber\\  \quad C_1 &=& {2
\sqrt{2}\zeta(3)\over\pi^2}\simeq 0.345, \quad C_2^{\alpha} = {2
\sqrt{2}\zeta(3)\over\pi^2} {A_{\alpha}\over G_F M^2_W}\simeq
0.5846 \, (0.1623),
\end{eqnarray}
for  $\alpha=e\,(\mu,\tau)$, and $E$ is the neutrino energy. The
quantity
\begin{equation}
\eta \doteq L^{(\alpha)}
\end{equation}
is the so-called effective asymmetry (abbreviated to $L$ in the
body of this Comment).

The neutrino matrix and its expansion in terms of the Pauli
matrices are
\begin{equation}\label{eq:dm}
\rho(p)= \left(
\begin{array}{cc}
\rho_{aa} & \rho_{a s} \\ \rho_{sa} & \rho_{ss}
\end{array}\right)
\doteq {f^0_{\rm eq}\over 2}\left(
\begin{array}{cc}
P_0+P_z & P_x-iP_y \\ P_x+iP_y & P_0-P_z \end{array}\right).
\end{equation}
 For the four matrix elements, DHPS
write down the following kinetic equations:
\begin{eqnarray}
\label{dotrhoaa} i\dot{\rho}_{aa} &=& F_0(\rho_{sa}-\rho_{as})/2
-i \Gamma_0 (\rho_{aa}-f^0_{eq}), \nonumber  \\ i \dot{\rho}_{ss}
&=& -F_0(\rho_{sa}-\rho_{as})/2, \nonumber \\ i \dot{\rho}_{as}
&=& W_0\rho_{as} +F_0(\rho_{ss}-\rho_{aa})/2- i \Gamma_1
\rho_{as}, \nonumber \\ i \dot{\rho}_{sa} &=& -W_0\rho_{sa} -
F_0(\rho_{ss}-\rho_{aa})/2- i\Gamma_1 \rho_{sa},
\end{eqnarray}
with the quantities $W_0$ and $F_0$ given by the expressions
\begin{equation}
F_0={\delta m^2 \over 2p}\sin 2\theta_0 \doteq V_x, \quad
W_0={\delta m^2 \over 2p}\cos 2\theta_0+ V^{a}_{eff} \doteq -V_z,
\end{equation}
and we have used $E \simeq p$.  The rates $\Gamma_0$ and
$\Gamma_1$ for $\alpha=e\,(\mu,\tau)$, and their translations are
\begin{eqnarray}\label{gammaj1}
\Gamma_0 &=& 2\Gamma_1  = g_a  {180 \zeta(3)\over 7 \pi^4} G_F^2
T^5 {y \over 3.15}  = 1.13\,(0.794)  G_F^2 T^5 y, \nonumber \\
\Gamma_0 & \doteq & \Gamma, \quad {\Gamma_1} \doteq D.
\end{eqnarray}
However, the coefficient $g_a/3.15$ is smaller than $k_{\alpha}$
[see Eq.\ (\ref{eq:gamma})]  found in  Ref.\ \cite{ekt}. This
discrepancy simply leads to a different correspondence of the
final asymmetry with a given choice of parameters $(\sin^2
2\theta_0,\delta m^2)$, and is thus irrelevant for the solution of
the controversy.

Are the two sets of kinetic equations (\ref{eq:b1}) and
(\ref{dotrhoaa}) equivalent? Rewriting the former in terms of
$P_{x,y,z,0}$ by way of Eq.\ (\ref{eq:dm}), one finds that they
are indeed concordant, save for the
 replacement $V_z\rightarrow -V_z$ which is just a matter of
convention. {\it Thus we conclude that approximation schemes in
DHPS and in Refs.\ \cite{ftv,fv1,bvw} begin with identical QKEs.}
However, as discussed in the main text, DHPS use the Fermi--Dirac
distribution with zero chemical potential as the active neutrino
equilibrium distribution function.

Moving on, DHPS replace the variables $\rho_{ij}$ with ones
closely connected to $P_0$ and ${\bf P}$:
\begin{eqnarray}
\rho_{\alpha\alpha}\equiv f_{\rm eq}^{0}[1+a] &\Rightarrow&
a\doteq \frac{1}{2}(P_0+P_z)-1=z_{\alpha}-1, \nonumber
\\ \rho_{s s} \equiv f_{\rm eq}^{0}[1+s] &\Rightarrow& s\doteq
\frac{1}{2}(P_0-P_z)-1=z_{s}-1, \nonumber
\\ \rho_{\alpha s}\equiv f_{\rm eq}^{0}[h+i\,l] &\Rightarrow& h\doteq
\frac{P_x}{2},\quad l\doteq -\frac{P_y}{2}.
\end{eqnarray}
They then take sums and differences of corresponding functions for
neutrinos and antineutrinos, i.e., for any generic variable $X$
and its antineutrino counterpart $\overline{X}$, they  define
\begin{equation}
X^{\pm}\equiv \frac{X\pm \overline{X}}{2}.
\end{equation}

The time integration variable is changed  to the inverse of the
dimensionless temperature
\begin{equation}
x\equiv \frac{\rm MeV}{T},
\end{equation}
which generates a factor ${\cal H} x$. Therefore DHPS  introduce
the following new quantities:
\begin{eqnarray}\label{eq:gamma2}
\gamma &=&{\Gamma_1\over {\cal H}x}\doteq \frac{D}{{\cal H}x},
\nonumber
\\ F &=& {F_0\over {\cal H} x}={\delta m^2 \over 2p {\cal H} x}\sin 2\theta_0
\doteq {V_x\over {\cal H} x},\nonumber \\ W &=& {W_0\over {\cal
H}x}={\delta m^2 \over 2p {\cal H} x}\cos
2\theta_0+\frac{V^{a}_{eff}}{{\cal H} x} \doteq -{V_z\over {\cal
H}x}.
\end{eqnarray}
They also find it useful to separate $W$ in two different
contributions, $W=U \pm VZ$, with
\begin{eqnarray}
\label{eq:U} U&=&{\delta m^2 \over 2p{\cal H}x} \left(\cos
2\theta_0 + C_2^a \frac{G^2_F T^4 E}{\alpha}\right) \doteq {\delta
m^2 \over 2p{\cal H}x}(c-b), \nonumber
\\ VZ&=&{C_1 \eta G_FT^3\over {\cal H}x}\doteq -{\delta m^2 \over
2p{\cal H}x}a,
\end{eqnarray}
where
\begin{equation}
 V = {29.6\over x^2}, \quad Z = 10^{10}\left (\eta_{0} - \int^{\infty}_0
  \frac{dy}{4 \pi^2} y^2 f^0_{\rm
eq} a_- \right).
\end{equation}
The quantity $Z$, as defined above, is clearly connected to the
lepton asymmetry $\eta$, although their exact relationship is not
specified by DHPS. This we provide here for completeness:
\begin{equation}
\eta = {4 \pi^2\over\zeta(3)}\,Z.
\end{equation}
%However, DHPS adopt $\eta=\pi^2 Z$, i.e., they are re-defining
%$\eta$, which must now be compared with $\zeta(3) L/4$, instead of
%$L$: the controversy is alleviated by a factor $\zeta(3)/4$. But
%this is clearly just a minor algebraic slip or a typographical
%error so the controversy remains.

 With
these definitions, DHPS rewrite the QKEs in the new variables
$s_{\pm}$, $a_{\pm}$, $h_{\pm}$, $l_{\pm}$:
\begin{eqnarray}
 \label{firstasyms}
 s_{\pm}' &=& F l_{\pm}, \nonumber \\
a_{\pm}' &=& - F l_{\pm} - 2 \gamma_+ a_{\pm} - 2 \gamma_-
a_{\mp}, \nonumber \\ h_{\pm}' &=& U l_{\pm} - V Z l_{\mp} -
\gamma_+ h_{\pm} - \gamma_- h_{\mp}, \nonumber \\ l_{\pm}'
&=&\frac{F}{2}(a_{\pm} - s_{\pm}) - U h_{\pm} + V Z h_{\mp} -
\gamma _+ l _{\pm} - \gamma_- l_{\mp},
\end{eqnarray}
in which terms containing
$\gamma_{-}=(\Gamma_1-\overline{\Gamma}_1)/2$ are later discarded.
This is also the case in Eq.\ (\ref{eq:b1}). Thus we stress again
that the DHPS kinetic equations are equivalent to Eq.\
(\ref{eq:b1}).

Now DHPS begin to extract their evolution equations by seeking
approximate solutions to Eq.\ (\ref{firstasyms}). More new
notations and symbols are introduced during the derivation, in
particular, the integration variable $x$ is replaced with $\tau$
by way of the function $U$:
\begin{equation}
U=\cos 2\theta_0 {\delta m^2 \over 2p{\cal H} x}
\left({1\over\tau^2}-1\right),
\end{equation}
such that $U=0$ at $\tau =1$.  Then it follows that
\begin{equation}
\tau=\xi {x^3\over y}\doteq \sqrt{c\over b}=\left(\frac{T_{\rm
res}^0}{T}\right)^3,
\end{equation}
where $\xi\simeq 6.5\,(12.8)\times 10^3 \sqrt{\cos 2\theta_0
{|\delta m^2|\over {\rm eV}^2}}$, for $\alpha = e\, (\mu,\tau)$.

At this point DHPS divide all the equations
by the quantity $M$, where
\begin{equation}
M\doteq {\delta m^2 c \over 2p{\cal H} x} \simeq 1.12\times 10^9
{x^2\over y} {|\delta m^2|\over {\rm eV}^2},
\end{equation}
and make the replacements ${1\over
M}(\gamma,F,U,VZ)\longrightarrow (\gamma,F,U,VZ)$, leading to
\begin{eqnarray}
&& \gamma={2p\over c\delta m^2} \Gamma_1 ={\delta\over\tau^2}
\doteq c d \simeq d, \nonumber \\ && F=-\tan 2\theta\simeq - s,
\nonumber \\ && U={1\over \tau^2} -1 \doteq {b\over
c}-1=\left(\frac{T}{T_{\rm res}^0}\right)^6-1, \nonumber \\&&
VZ\doteq {a\over c}\simeq a.
\end{eqnarray}
These new definitions combine to produce the change
\begin{equation}
{d\over dx}\longrightarrow {1\over Q}{d\over d\tau},
\end{equation}
in the left hand side of the kinetic equations, where
\begin{equation}
Q = M \left({d\tau\over dx}\right)^{-1} \simeq 5.7\times 10^4
\sqrt{{|\delta m^2|\over {\rm eV}^2} \cos 2\theta_0}.
\end{equation}

More new definitions follow from here, whose translations are
listed below (with $c\simeq 1$):
\begin{eqnarray}
D&=&ZV\doteq a, \nonumber \\ \sigma^2&=&\gamma^2+U^2\doteq
d^2+(c-b)^2, \nonumber \\ \tilde{U}&=&U \left(1-{D^2\over\sigma^2}
\right)\doteq (b-1)\left[1- \frac{a^2}{d^2+(1-b)^2}\right],
\nonumber \\ \tilde{\gamma}&=&\gamma
\left(1+{D^2\over\sigma^2}\right) \doteq d
\left[1+\frac{a^2}{d^2+(1-b)^2}\right], \nonumber \\
\tilde{\sigma}&=&\tilde{\gamma}^2+\tilde{U}^2, \nonumber \\
t&=&{1\over\tau}\doteq \left(\frac{T}{T_{\rm res}^0}\right)^3=
\sqrt{b\over c}\simeq \sqrt{b}.
\end{eqnarray}
The quantity $\tau$ is inversely proportional to $y$. DHPS thus
define the new variable
\begin{equation}
q=y\tau \doteq \left({T_{\rm res}^0|_{y=1}\over T}\right)^3.
\end{equation}
Moreover they introduce a new function $b_0$ that expresses the
difference between sum distribution of active neutrinos and sum
distribution of sterile neutrinos:
\begin{equation}
b_0\simeq -s^+\doteq 1-z^{+}_s,
\end{equation}
such that their final sterile neutrino production equation [Eq.\
(42) in DHPS] is
\begin{equation}\label{eq:sp}
{db_0\over d\tau} = - {Q F^2 \tilde\gamma \over 2\tilde\sigma^2} b_0,
\end{equation}
while the asymmetry evolution equation [their Eq.\ (51)] is given
by
\begin{equation}\label{eq:dL}
\frac{1}{Z}\frac{d Z}{dq} = - \delta B   q^{5/3} \int_0^\infty dt
 { t^4 (t^2-1) f_{eq} (tq) b_0(1/t) \over \sigma^2 \tilde\sigma^2
} \left[ 1-{F^2 ( \sigma^2 +D^2) \over 4 \sigma^2 \tilde\sigma^2 }
\right].
\end{equation}

With some trivial algebra one can show that
\begin{eqnarray}
\sigma^2\tilde{\sigma}^2&=&[\gamma^2+(U+D)^2] [\gamma^2+(U-D)^2],
\nonumber \\ \sigma^2+D^2&=&\gamma^2+U^2+D^2.
\end{eqnarray}
Using the conversion relations, the DHPS sterile neutrino
production equation translates to
\begin{equation}
{\partial z^+_s\over \partial (T/{\rm MeV})}=-{s^2\Gamma\over 4
{\cal H}T} {d^2+(b-c)^2+a^2 \over \Delta(0)} [1-z^+_s],
\end{equation}
 while their neutrino asymmetry evolution equation
is equivalent to
\begin{equation}
{1\over L} \frac{dL}{d(T/{\rm MeV})}=\int_0^{\infty}
{A\over\Delta(0)} \left[1-{s^2\over 4}
\frac{d^2+(c-b)^2+a^2}{\Delta(0)} \right] dy,
\end{equation}
in our language.

\vskip 0.5cm 
\noindent
{\bf Note Added}
\vskip 0.5cm
\noindent
Shortly after completion of this paper, R. Buras and D. V.
Semikoz informed us of their work (hep-ph/0008263),
where they also concluded that the main result of
DHPS was incorrect.

\begin{figure}
\caption{Evolution of the effective total lepton number,
$L^{(\tau)}$, for $\nu_{\tau} \leftrightarrow \nu_s$ oscillations
for the parameter choice, $\delta m^2 = -10\, {\rm eV}^2$, $\sin^2
2\theta_0 = 10^{-7}$. In Fig.\ 1a, the solid curve is the
numerical solution of the QKEs while the dashed curve is the
static approximation for $r|_{\rm res} \equiv (\ell_{\rm
int}/\ell_{m})|_{\rm res} < 1$ while for $r|_{\rm res} > 1$ pure
adiabatic MSW evolution is employed. The dashed dotted line is the
static approximation (over all T). Fig.\ 1b contains the
corresponding evolution of the MSW resonance momentum/T (i.e. $y_{\rm
res}$) for the neutrinos (solid line) and anti-neutrinos (dashed
line) for this example.}

\vspace{10mm}

\caption{Same as Fig.\ 1 except for the parameter choice $\delta
m^2 = -100\, {\rm eV}^2$ and $\sin^2 2\theta_0 = 10^{-8}$.}

\newpage
\begin{center}
\epsfig{file=dom1.eps,width=15cm}

\newpage
\epsfig{file=dom1b.eps,width=15cm}
\newpage
\epsfig{file=dom2.eps,width=15cm}
\newpage
\epsfig{file=dom2b.eps,width=15cm}

\end{center}
\end{figure}

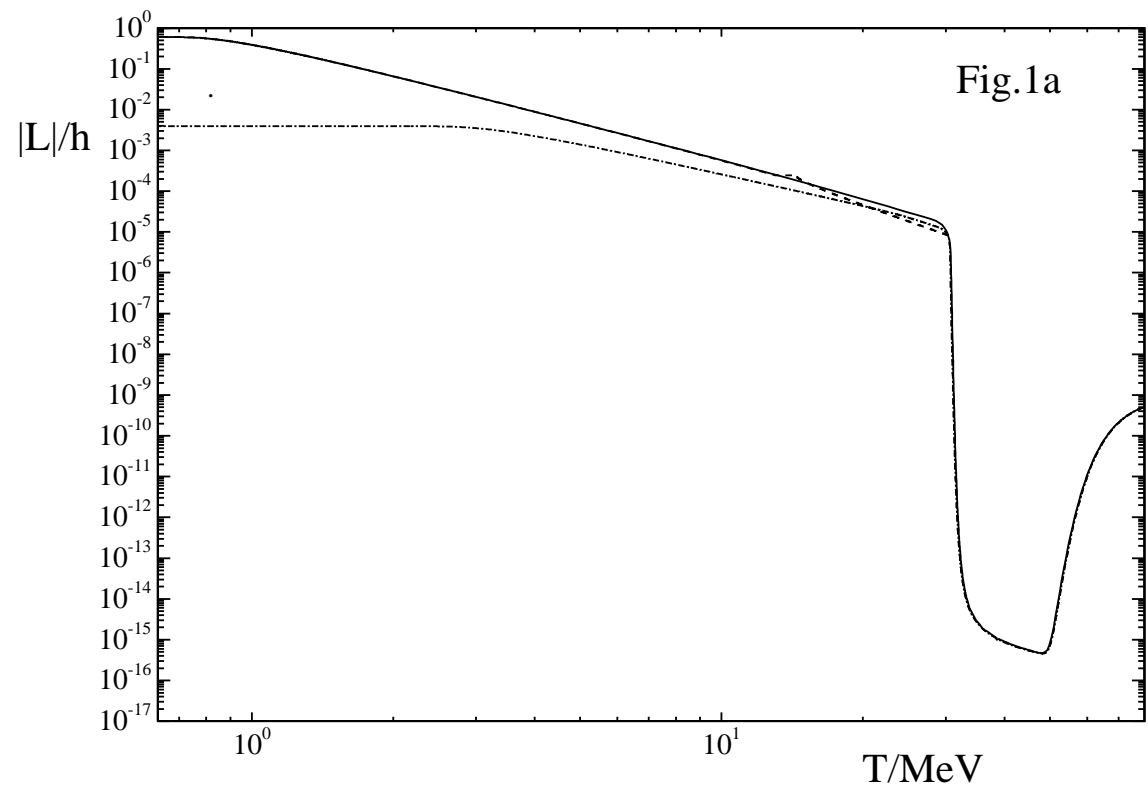
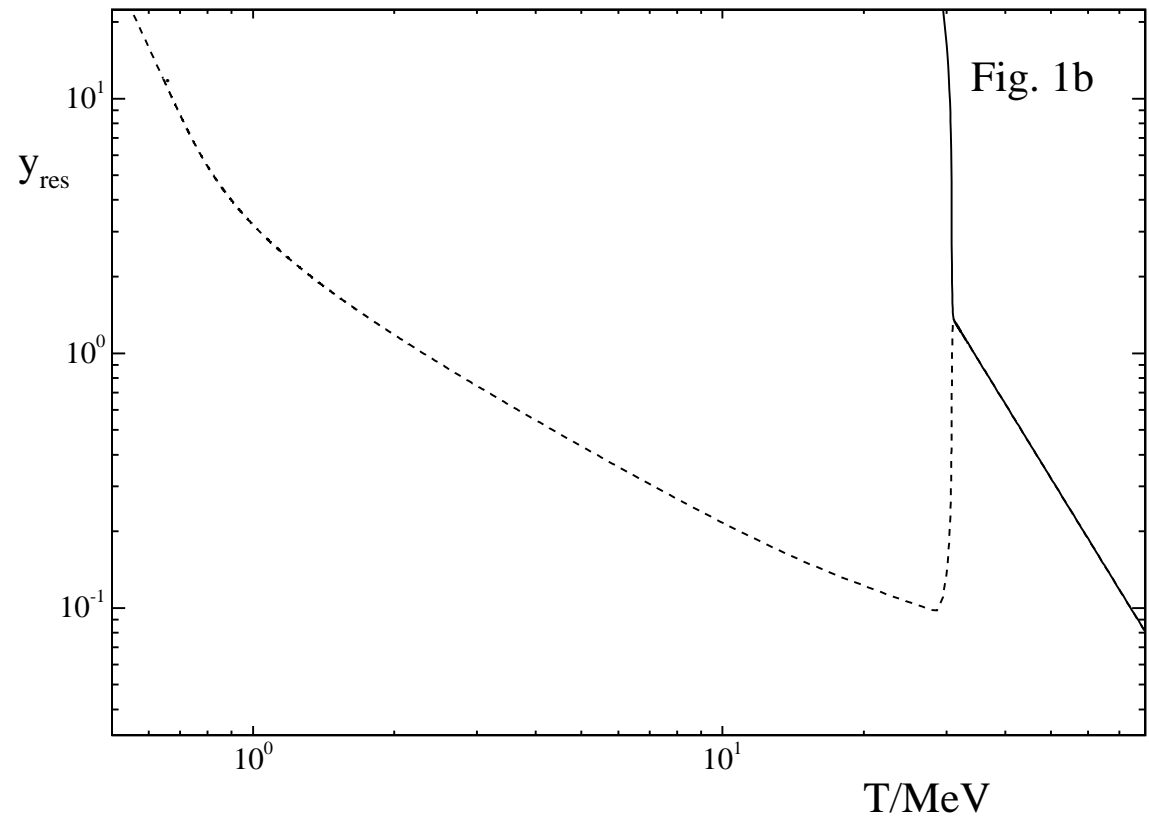
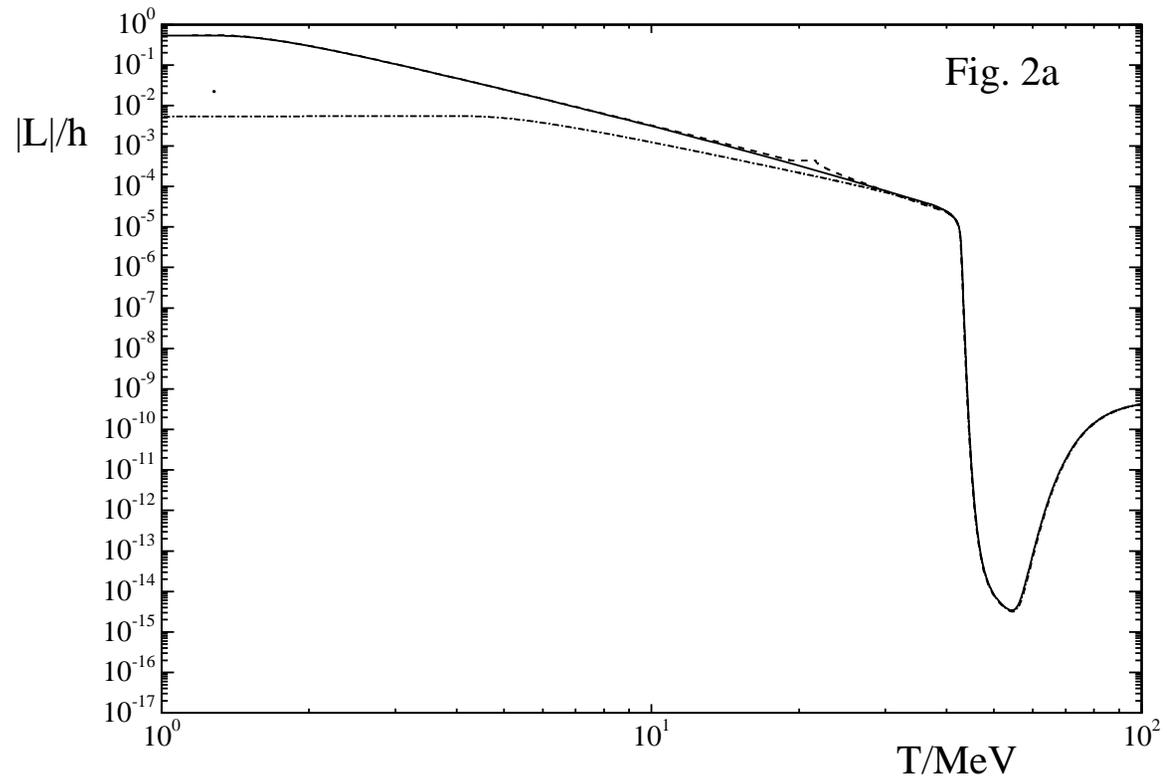
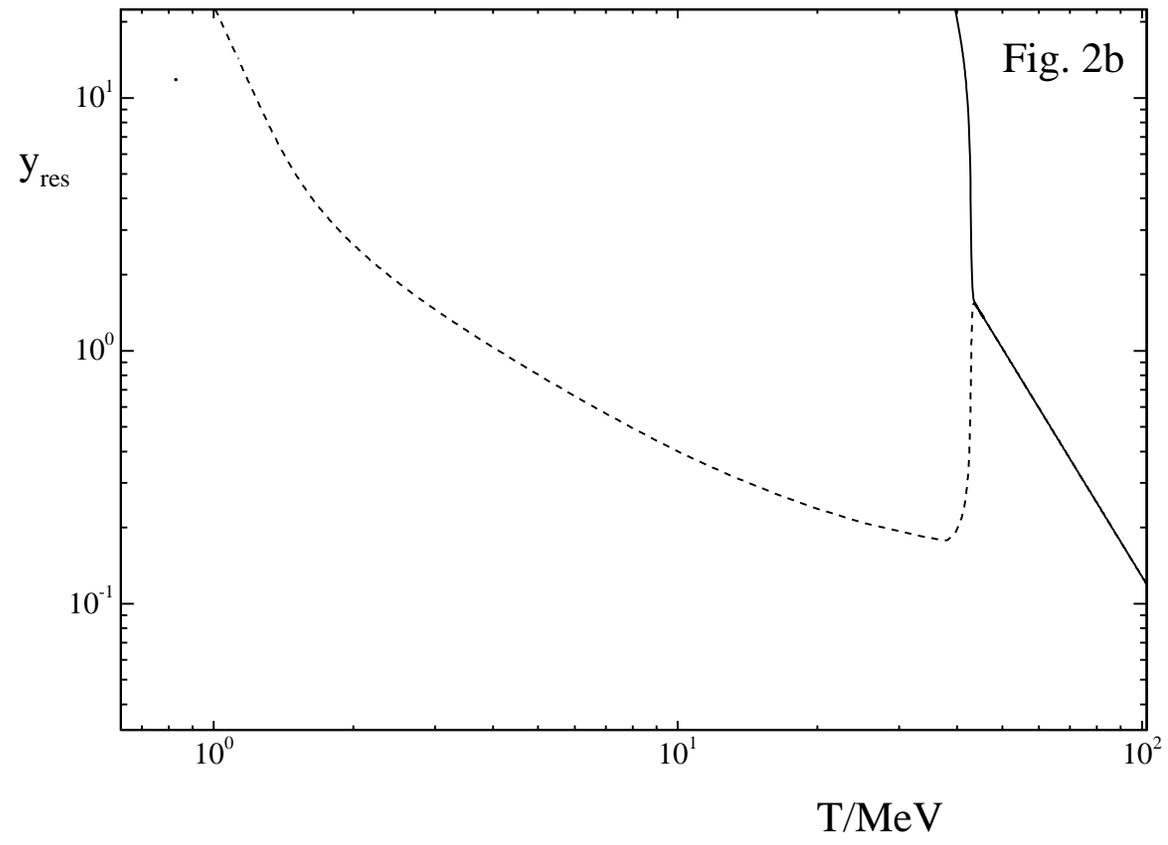
\end{document}